\documentclass[MSNbibl,dvips]{arxstspdf}
\usepackage{graphicx}
\usepackage{flushend}
\usepackage{stfloats}

\volume{26}
\issue{3}
\pubyear{2011}
\firstpage{423}
\lastpage{439}
\doi{10.1214/11-STS357}

\makeatletter
\newproclaim{definition}{Definition}
\newtheorem{lemma}{Lemma}
\newtheorem{theorem}{Theorem}
\newtheorem{alemma}{Lemma}[section]
\newproclaim{example}{Example}
\newcommand{\eqref}[1]{(\ref{#1})}
\makeatother

\begin{document}
\begin{frontmatter}

\title{Weak Informativity and the Information in One Prior Relative to Another}
\runtitle{Weak Informativity}

\begin{aug}
\author{\fnms{Michael} \snm{Evans}\corref{}\ead[label=e1]{mevans@utstat.utoronto.ca}}
and
\author{\fnms{Gun Ho} \snm{Jang}\ead[label=e2]{gunjang@upenn.edu}}
\runauthor{M. Evans and G. H. Jang}

\address{Michael Evans is Professor,
Department of Statistics,
University of Toronto,
Toronto, Ontario, M5S 3G3, Canada
\printead{e1}.
Gun Ho Jang is Postdoctoral Fellow,
Department of Biostatistics and Epidemiology,
University of Pennsylvania,
Philadelphia, Pennsylvania 19104, USA
\printead{e2}.}

\affiliation{University of Toronto and University of Pennsylvania}

\end{aug}

% ABSTRACT
%
\begin{abstract}
A question of some interest is how to characterize the amount of information
that a prior puts into a statistical analysis. Rather than a general
characterization, we provide an approach to
characterizing the amount of information a prior puts into an analysis, when
compared to another base prior. The base prior is considered to be the prior
that best reflects the current available information. Our purpose then is
to characterize priors that can be used as conservative inputs to an
analysis relative to the base prior. The characterization that we
provide is in
terms of a priori measures of prior-data conflict.
\end{abstract}

% KEYWORDS
%

\end{frontmatter}

%s1 ###
\section{Introduction}\vspace*{3pt}\label{sec:01}

Suppose we have two proper priors $\Pi_{1}$ and $\Pi_{2}$ on a parameter
space $\Theta$ for a statistical model $\{P_{\theta}\dvtx \theta\in
\Theta\}$.
A natural question to ask is: how do we compare the amount of information
each of these priors puts into the problem? While there may seem to be
natural intuitive ways to express this, such as prior variances, it seems
difficult to characterize this precisely in general. For example, the
consideration of several examples in Sections \ref{sec:03} and \ref
{sec:04} makes it clear that
using the variance of the prior is not appropriate for this task.

The motivation for this work comes from Gelman (\citeyear
{Gelman2006}) and Gelman et al. (\citeyear{Gelman-etal2008}),
where the intuitively satisfying notion of weakly informative
priors is introduced as a compromise between informative and noninformative
priors. The basic idea is that we have a base prior $\Pi_{1}$, perhaps
elicited, that we believe reflects our current information about
$\theta$,
but we choose to be conservative in our inferences and select a prior
$\Pi
_{2}$ that puts less information into the analysis. While it is common to
take $\Pi_{2}$ to be a noninformative prior, this can often produce
difficulties when $\Pi_{2}$ is improper, and even when $\Pi_{2}$ is
proper, it seems inappropriate, as it completely discards the
information we
have about $\theta$ as expressed in $\Pi_{1}$. In addition,
we may find that a prior-data conflict exists with $\Pi_{1}$ and so
look for another
prior that reflects at least some of the information that $\Pi_{1}$
puts into an
analysis, but avoids the conflict.

We note that our discussion here is only about how we should choose
$\Pi_{2}
$ \textit{given} that $\Pi_{1}$ has already been chosen. Of course, the
choice of $\Pi_{1}$ is of central importance in a Bayesian analysis.
Ideally, $\Pi_{1}$ is chosen based on a clearly justified elicitation
process, but we know that this is often not the case. In such a circumstance
it makes sense to try and choose $\Pi_{1}$ reasonably but then be
deliberately less informative by choosing $\Pi_{2}$ to be weakly
informative with respect to $\Pi_{1}$. The point is to inspire confidence
that our analysis is not highly dependent on information that may be
unreliable. To do this, however, requires a definition of what it means for
one prior to be weakly informative with respect to another and that is what
this paper is about.

To implement the idea of weak informativity, we need a precise
definition. We provide this in Section~\ref{sec:02}
and note that it involves the notion of prior-data conflict.
Intuitively, a prior-data
conflict occurs when the prior places the bulk of its mass where the
likelihood is relatively low, as the likelihood is indicating that the
true value of the parameter
is in the tails of the prior. Our definition of weak
informativity is then expressed by saying that $\Pi_{2}$ is weakly
informative relative to $\Pi_{1}$ whenever $\Pi_{2}$ produces fewer
prior-data conflicts a priori than $\Pi_{1}$. This leads to a
quantifiable
expression of weak informativity that can be used to choose priors. In
Section \ref{sec:03} we consider this definition in the context of
several standard
families of priors and it is seen to produce results that are intuitively
reasonable. In Section~\ref{sec:04} we consider applications of this
concept in some
data analysis problems. While our intuition about weak informativity is
often borne out, we also find that in certain situations we have to be
careful before calling a prior weakly informative.

First, however, we establish some notation and then review how we check for
prior-data conflict. We suppose that $P_{\theta}(A)=\int_{A}f_{\theta
}(x)\mu(dx)$, that is, each $P_{\theta}$ is absolutely continuous with respect
to a support measure $\mu$ on the sample space $\mathcal{X}$, with the
density denoted by $f_{\theta}$. With this formulation a prior $\Pi$ leads
to a prior predictive probability measure on $\mathcal{X}$\ given by $%
M(A)=\int_{\Theta}P_{\theta}(A) \Pi(d\theta)=\int_{A}m(x) \mu(dx)$,
where $m(x)=\int_{\Theta}f_{\theta}(x) \Pi(d\theta)$. If $T$ is a
minimal sufficient statistic for $\{P_{\theta}\dvtx \theta\in\Theta\}$, then
it is well known that the posterior is the same whether we observe $x$
or~$T(x)$. So we will denote the posterior by $\Pi(\cdot| T)$ hereafter.
Since $T$ is minimal sufficient, we know that the conditional
distribution of
$x$ given $T$ is independent of $\theta$. We denote this conditional
measure by $P(\cdot| T)$. The joint distribution $P_{\theta}\times
\Pi$
can then be factored as
%
%e1 ###
\begin{eqnarray}\label{eq:01}
P_{\theta}\times\Pi&=&M\times\Pi(\cdot| x)
\nonumber
\\[-8pt]
\\[-8pt]
\nonumber
&=&P(\cdot| T)\times
M_{T}\times\Pi(\cdot| T),
\end{eqnarray}
where $M_{T}$ is the marginal prior predictive distribution of $T$.

While much of Bayesian analysis focuses on the third factor in \eqref
{eq:01}, there
are also roles in a statistical analysis for $P(\cdot| T)$ and $M_{T}$.
As discussed in Evans and Moshonov (\citeyear{EvansMoshonov2006,EvansMoshonov2007}), $P(\cdot| T)$ is
available for checking the sampling model, for example, if $x$ is a surprising
value from this distribution, then we have evidence that the model $%
\{P_{\theta}\dvtx \theta\in\Theta\}$ is incorrect. Furthermore, it is
argued that,
if we conclude that we have no evidence against the model, then the
factor $%
M_{T}$ is available for checking whether or not there is any prior-data
conflict, and we do this by comparing the observed value of $T(x)$ to $M_T$.
If we have no
evidence against the model, and no evidence of prior-data conflict,
then we
can proceed to inferences about $\theta$.
Actually, the issues involved in model checking and checking for prior-data
conflict are more involved than this (see, e.g., the cited references
and Section \ref{sec:05}), but \eqref{eq:01} gives the basic idea that
the full information, as
expressed by the joint distribution of $(\theta,x)$, splits into
components, each of which is available for a specific purpose in a~statistical analysis.

Accordingly, we restrict ourselves here, for any discussions concerning
prior-data conflict,
to working with $M_{T}$. One issue that needs
to be addressed is how one is to compare the observed value
$t_{0}=T(x_{0})$ to $M_{T}$. In essence, we need
a measure of surprise and for this we use a $P$-value.
Effectively, we are in the situation where we have a value from a single
fixed distribution and we need to specify the appropriate $P$-value to
use. In
Evans and Moshonov (\citeyear{EvansMoshonov2006,EvansMoshonov2007}) the $P$-value for checking for prior-data
conflict is given by
%
%e2 ###
\begin{equation}
M_{T}\bigl(m_{T}(t)\leq m_{T}(t_{0})\bigr), \label{eq:02}
\end{equation}
where $m_{T}$ is the density of $M_{T}$ with respect to the volume
measure on the range
space for $T$.
In Evans and Jang (\citeyear{EvansJang2010b}) it is proved that, for many of the models and
priors used in statistical
analyses, \eqref{eq:02} converges almost surely, as the amount of data
increases, to $\Pi(\pi(\theta) \leq\pi(\theta_*))$, where
$\theta_*$ is the true value of $\theta$. So \eqref{eq:02} is assessing
to what extent
the true value is in the tails of the prior, or, equivalently, to what
extent the prior information
is in conflict with how the data is being generated.

A difficulty with \eqref{eq:02} is that it is not generally invariant
to the choice of
the minimal sufficient statistic $T$. A general invariant $P$-value is
developed in Evans and Jang (\citeyear{EvansJang2010annstat}) for situations
where we want to compare the observed value of a statistic to a~fixed
distribution.
This requires that the model and $T$ satisfy some regularity
conditions, for example, all spaces
need to be locally Euclidean, support measures are given by volume
measures on these spaces,
and~$T$ needs to be sufficiently smooth. A formal
description of these conditions can be found in Tjur (\citeyear{Tjur1974})
and it is noted that these hold for the typical statistical application.
For example,
these conditions are immediately satisfied in the discrete case.
Furthermore, for continuous
situations,\vadjust{\goodbreak} with densities defined as limits, we get the usual
expressions for densities.
When applied to checking for prior-data conflict,
this leads to using the invariant $P$-value
%
%e3 ###
\begin{equation}
M_{T}\bigl(m_{T}^{\ast}(t)\leq m_{T}^{\ast}(t_{0})\bigr), \label{eq:03}
\end{equation}
where $m_{T}^{\ast}(t)= \int_{T^{-1}{t} } m(x) \mu_{T^{-1}\{t\}
}(dx)= m_{T}(t)\cdot\break  E(J_T^{-1}(x) | T(x)=t)$,
$\mu_{T^{-1}\{t\}}$ is the volume measure on $T^{-1}\{t\}$,
$J_{T}(x)=(\det(dT(x)\circ dT^{\prime
}(x)))^{-1/2}$ and $dT$ is the differential of $T$.
Note that $J_{T}(x)$ gives the volume distortion produced by $T$ at $x$.
So $m_{T}^{\ast}$ is the density of $M_{T}$ with respect to the
support measure
given by $\{E(J_T^{-1}(x) | T(x)=t)\}^{-1}$ times the volume
measure on the range space for $T$.

In applications all models are effectively discrete, as we measure
responses to some finite accuracy,
and continuous models are viewed as being approximations. The
use of \eqref{eq:03}, rather than \eqref{eq:02}, then expresses the
fact that we do not want volume distortions induced by a transformation
to affect
our inferences. So we allocate this effect of the transformation with
the support measure, rather than with the density,
when computing the $P$-value.
In the discrete case, as well as when $T$ is linear, \eqref{eq:02} and
\eqref{eq:03} give the same value and otherwise
seem to give very similar values.
Convergence of \eqref{eq:03}, to an invariant $P$-value based on the
prior, is established in Evans and Jang (\citeyear{EvansJang2010b}). We use \eqref{eq:03}
throughout this paper but note that it is only in Section \ref
{sec:03.3} where \eqref{eq:03} differs from~\eqref{eq:02}.

Our discussion here is based on a minimal sufficient statistic $T$. We
note that, except in mathematically
pathological situations, such a statistic exists. It may be, however,
that $T$ is high dimensional, for example,
$T$ can be of the same dimension as the data. In such situations the
dimensionality of the problem can often be reduced
by examining
components of the prior in a hierarchical fashion. For example, when
the prior on $\theta
=(\theta_{1},\theta_{2})$ is specified as $\pi(\theta)=\pi
_{2}(\theta
_{2} | \theta_{1})\pi_{1}(\theta_{1})$, then $\pi_{1}$ and $\pi
_{2}(\cdot| \theta_{1})$ are checked separately and so the definition
of weak informativity applies to each component separately. This is
exemplified by the regression example of Section \ref{sec:04.2} where
$\theta=(\theta
_{1},\theta_{2})=(\beta,\sigma^{2})$. More on checking the
components of
a prior can be found in Evans and Moshonov (\citeyear
{EvansMoshonov2006}). Furthermore, when ancillaries
exist, it is necessary to condition on these when checking for
prior-data conflict,
as this variation has nothing to do with the prior. This results in a
reduction of the dimension of the
problem. The relevance of ancillarity to the problem of weakly
informative priors is discussed in Section \ref{sec:05}.

When choosing a prior it makes sense to consider the prior distribution
of more than just the minimal
sufficient statistic. For example, Chib and Ergashev (\citeyear
{ChibErgashev2009}) consider the prior distribution of a~somewhat
complicated function of the
parameters and data that has a real world interpretation. If this
distribution produces values that seem reasonable
in light of what is known, then this goes some distance toward
justifying the prior. Also, the level of informativity of the
prior can be judged by looking at the prior distribution of this
quantity when that is possible. While this is certainly a
reasonable approach to choosing $\Pi_{1}$, it does not supply us with
a definition of weak informativity.
For example, a prior $\Pi_{1}$ can be chosen as discussed in Chib and
Ergashev (\citeyear{ChibErgashev2009}), but then $\Pi_{2}$ could be
chosen to be weakly informative with respect to
$\Pi_{1}$, to inspire confidence that conclusions drawn are not highly
dependent on subjective appraisals.

As we will show, there will typically be many priors $\Pi_{2}$ that
are weakly informative with respect to a~given base prior $\Pi_{1}$.
The question then arises as to which $\Pi_{2}$ we should use. This is
partially answered in Section \ref{sec:02}
where we show that the definition of weak informativity leads to a
quantification of how much less informative
$\Pi_{2}$ is than $\Pi_{1}$. For example, we can choose $\Pi_{2}$ in
a family of priors to be 50\% less informative than $\Pi_{1}$.
Still, there may be many such $\Pi_{2}$ and at this time we do not
have a criterion that allows us to distinguish among such priors. For example,
suppose the base prior is a normal prior for a~location parameter. We
can derive weakly informative priors with respect to such a prior in
the family of normal priors
(see Section \ref{sec:03.1}) or in the family of~$t$ priors (see
Section \ref{sec:03.2}). There is nothing in our developments that
suggests that a weakly informative~$t$ prior is to be preferred to a
weakly informative normal prior or conversely. Such distinctions will
have to be made based on other criteria.

%s2 ###
\section{Comparing Priors}\label{sec:02}

There are a variety of measures of information~used in statistics. Several
measures have been based on~the concept of entropy, for example, see
Lindley (\citeyear{Lindley1956})
and Bernardo (\citeyear{Bernardo1979}). While these measures have
their virtues, we note that
their coding theory interpretations can seem somewhat abstract in
statistical contexts and they can suffer from nonexistence in certain
problems. Also, Kass and Wasserman (\citeyear{KassWasserman1995})
contain some discussion concerned
with expressing the absolute information content of a prior in terms of
additional sample values. Rather than adopting these approaches, we consider
comparing priors ba\-sed on their tendencies to produce prior-data
conflicts. This formulation of the relative amount of information put
into an
analysis has a direct interpretation in terms of statistical consequences.

Suppose that an analyst has in mind a prior $\Pi_{1}$ that they believe
represents the information at hand concerning $\theta$. The analyst,
however, prefers to use a~prior $\Pi_{2}$\ that is conservative,
when compared to~$\Pi_{1}$. In such a situation it seems reasonable to
consider~$\Pi_{1}$ as a base prior and then compare all other priors to it. This idea
comes from Gelman (\citeyear{Gelman2006}) and leads to the notion of
weakly informative
priors.

Before we observe data we have no way of knowing
if we will have a prior-data conflict. Accordingly, since the analyst has
determined that $\Pi_{1}$ best reflects the available information, it is
reasonable to consider the prior distribution of
$P_{1}(t_{0})=\break M_{1T}(m_{1T}^{\ast}(t)\leq m_{1T}^{\ast}(t_{0}))$ when
$t_{0}\sim M_{1T}$. Of course, this is effectively uniformly distributed
[exactly so when $m_{1T}^{\ast}(t)$ has a continuous distribution when
$%
t\sim M_{1T}$] and this expresses the fact that all the information about
assessing whether or not a prior-data conflict exists is contained in the
$P$-value, with no need to compare the $P$-value to its distribution.

$\!\!$Consider now, however, the distribution of
$P_{2}(t_{0})\!=M_{2T}(m_{2T}^{\ast}(t)\leq m_{2T}^{*}(t_{0}))$
which is used to check whe\-ther or not there
is prior-data conflict with respect to $\Pi_{2}$. Given that we have identified
that a priori the appropriate distribution of $t_{0}$ is $M_{1T}$,
at least for inferences about an unobserved value, then $P_{2}(t_{0})$ is
not uniformly distributed. In fact, from the distribution of $P_{2}(t_{0})$
we can obtain an intuitively reasonable idea of what it means for a
prior $%
\Pi_{2}$ to be weakly informative relative to $\Pi_{1}$. Suppose that
the prior distribution of $P_{2}(t_{0})$ clusters around 1. This implies
that, if we were to use $\Pi_{2}$ as the prior when $\Pi_{1}$ is
appropriate, then there is a small prior probability that a prior-data
conflict would arise. Similarly, if the prior distribution of $P_{2}(t_{0})$
clusters around 0, then there is a large prior probability that a prior-data
conflict would arise. If one prior distribution results in a larger prior
probability of there being a~prior-data conflict than another, then it seems
reasonable to say that the first prior is more informative than the second.
In fact, a completely noninformative prior should never produce prior-data
conflicts.

So we compare the distribution of $P_{2}(t_{0})$ when $t_{0}\sim
M_{1T}$, to
the distribution of $P_{1}(t_{0})$ when $t_{0}\sim M_{1T}$,\vadjust{\goodbreak} and do this
in a
way that is relevant to the prior probability of obtaining a prior-data
conflict. One approach to this comparison is to select a $\gamma
$-quantile $x_{\gamma}\in\lbrack0,1]$ of the distribution of $P_{1}(t_{0})$, and then
compute the probability
%
%e4 ###
\begin{equation}
M_{1T}\bigl(P_{2}(t_{0})\leq x_{\gamma}\bigr). \label{eq:04}
\end{equation}
The value $\gamma$ is presumably some cutoff, dependent on the
application, where we will consider that evidence of a prior-data conflict
exists whenever\break $P_{1}(t_{0})\leq\gamma$. Of course, if $m_{1T}^{\ast
}(t_{0})$ has a continuous distribution when $t_{0}\sim M_{1T}$, then $%
x_{\gamma}=\gamma$. Our basic criterion for the weak informativity of
$\Pi_{2}$
relative to~$\Pi_{1}$ will then be that \eqref{eq:04} is less than or
equal to~$x_{\gamma}$. This implies that the prior probability of obtaining
a~prior-data conflict under $\Pi_{2}$ is no greater than when $\Pi_{1}$ is
used, at least when we have identified~$\Pi_{1}$ as our correct
prior.

\begin{definition} If \eqref{eq:04} is less than or equal
to~$x_{\gamma}$, then $\Pi_{2}$ is \textit{weakly informative relative to} $\Pi_{1}$
\textit{at level}~$\gamma$. If $\Pi_{2}$ is weakly informative relative to
$\Pi_{1}$ at level $\gamma$ for every $\gamma\leq
\gamma_{0}$, then $\Pi_{2}$ is \textit{uniformly weakly informative relative
to} $\Pi_{1}$ \textit{at level}~$\gamma_{0}$. If $\Pi_{2}$ is weakly~in\-formative
relative to $\Pi_{1}$ at level $\gamma$ for every $\gamma$,
then~$\Pi_{2}$ is \textit{uniformly weakly informative relative to}~$\Pi_{1}$.
\end{definition}

Typically we would like to choose a prior $\Pi_{2}$ that is uniformly
weakly informative with respect to $\Pi_{1}$. This still requires us
to select a prior
from this class, however, and for this we must choose a level $\gamma$.

Once we have selected $\gamma$, the degree of weak informativity of a prior
$\Pi_{2}$ relative to $\Pi_{1}$ can be assessed by comparing $%
M_{1T}(P_{2}(t_{0})\leq x_{\gamma})$ to $x_{\gamma}$ via the ratio
%
%e5 ###
\begin{equation}
1-M_{1T}\bigl(P_{2}(t_{0})\leq x_{\gamma}\bigr)/x_{\gamma}. \label{eq:05}
\end{equation}
If $\Pi_{2}$ is weakly informative relative to $\Pi_{1}$\ at
level~$\gamma$, then \eqref{eq:05} tells us the proportion of fewer prior-data
conflicts we can
expect a priori when using $\Pi_{2}$ rather than $\Pi_{1}$. Thus,
\eqref{eq:05} provides a measure of how much less informative $\Pi
_{2}$ is than $\Pi
_{1}$ at level $\gamma$. So, for example, we might ask for a prior $\Pi
_{2}$ that is
uniformly weakly informative with respect to $\Pi_1$ and then, for a
particular $\gamma$,
select a prior in this class~such that~\eqref{eq:05} equals 50\%.

As we will see in the examples, it makes sense to talk of one prior being
\textit{asymptotically weakly informative at level }$\gamma$ with respect
to another prior in the sense that~\eqref{eq:04} is bounded above by
$\gamma$ in the
limit as the amount of data increases. In several cases this simplifies
matters considerably, as an asymptotically weakly informative prior is easy
to find and may still be weakly informative for finite amounts of data.

While \eqref{eq:04} seems difficult to work with, the following result
is proved in
the \hyperref[app]{Appendix} and gives a~simpler expression.% \smallskip

\begin{lemma}\label{lem:01}
Suppose $P_{i}(t)$ has a continuous distribution
under $M_{iT}$ for $i=1,2$. Then there exists~$r_{\gamma}$ such that $%
M_{1T}(P_{2}(t)\leq\gamma)=M_{1T}(m_{2T}^{\ast}(t)\leq r_{\gamma
})$, and $\Pi_{2}$ is weakly informative at level $\gamma$ relative to $\Pi
_{1} $ whenever $M_{1T}(m_{2T}^{\ast}(t) \leq r_{\gamma}) \leq\gamma$.
Furthermore, $\Pi_{2}$ is uniformly weakly informative relative to $\Pi
_{1}$ if and only if $M_{1T}(m_{2T}^{\ast}(t)\leq m_{2T}^{\ast
}(t_{0}))\leq M_{2T}(m_{2T}^{\ast}(t)\leq\break m_{2T}^{\ast}(t_{0}))$ for every
$t_{0}$.%\smallskip
\end{lemma}

Note that the equivalent condition for
uniform weak informativity in Lemma \ref{lem:01} says that the
probability content, under $M_{1T}$,
in the ``tails'' (regions of low density) of the density $m_{2T}^{\ast}$ is always
bounded above by the probability content under~$M_{2T}$. So~$M_{2T}$
puts more probability
content into these tails than $M_{1T}$ and this can be taken as an
indication that~$M_{2T}$
is more dispersed than $M_{1T}$. Lemma \ref{lem:01} typically applies
when we are dealing with continuous
distributions on $\mathcal{X}$. It can also be shown that~$P_{i}(t)$
has a continuous distribution under $M_{iT}$ if and only if~$m_{iT}^{\ast}(t)$
has a continuous distribution under~$M_{iT}$.

%s3 ###
\section{Deriving Weakly Informative Priors}\label{sec:03}

We consider several examples of families of priors that arise in
applications. These
examples support our definition of weak informativity and also lead to some
insights into choosing priors. The results obtained for the examples in
this section
are combined in Section~\ref{sec:04.2} to give results for a
practically meaningful context.

We first note that, while we could consider comparing arbitrary priors
$\Pi_2$ to
$\Pi_1$, we want $\Pi_2$ to reflect at least some of the information expressed
in $\Pi_1$. The simplest expression of this is to require that $\Pi_2$
have the same, or nearly the same,
location as $\Pi_1$. This restriction simplifies the analysis and seems natural.

%s3.1 ###
\subsection{Comparing Normal Priors}\label{sec:03.1}

Suppose we have a sample $x=(x_{1},\ldots,x_{n})$ from a~$N(\mu,1)$
distribution where $\mu$ is unknown. Then $t=T(x)=\bar{x}\sim N(\mu,1/n)$
is minimal sufficient and~sin\-ce~$T$ is linear, there is constant volume
distortion and so this can be ignored. Suppose that the
prior~$\Pi_{1}$ on~$\mu$ is a $N(\mu_{0},\sigma_{1}^{2})$ distribution with $\mu_{0}$
and $\sigma_{1}^{2}$ known. We then have that $M_{1T}$ is the $N(\mu
_{0},1/n+\sigma_{1}^{2})$ distribution. Now suppose that $\Pi_{2}$ is\vadjust{\goodbreak}
a $N(\mu_{0},\sigma_{2}^{2})$ distribution with $\sigma_{2}^{2}$ known. Then
$M_{2T}$ is the $N(\mu_{0},\break1/n+\sigma_{2}^{2})$ distribution and
\begin{eqnarray*}
P_{2}(t_{0})
&=& M_{2T}\bigl(m_{2T}^{\ast}(t)\leq m_{2T}^{\ast
}(t_{0})\bigr)\\
&=&M_{2T}\bigl(m_{2T}(t)\leq m_{2T}(t_{0})\bigr) \\
&=& M_{2T}\bigl((t-\mu_{0})^{2}\geq(t_{0}-\mu_{0})^{2}\bigr)\\
&=&1-G_{1}\bigl((t_{0}-\mu
_{0})^{2}/(1/n+\sigma_{2}^{2})\bigr),
\end{eqnarray*}
where $G_{k}$ denotes the $\operatorname{Chi\mbox{-}squared}(k)$ distribution function. Now
under $%
M_{1T}$ we have that $(t_{0}-\mu_{0})^{2}/\break (1/n+\sigma_{1}^{2})\sim
\operatorname{Chi\mbox{-}squared}(1)$. Therefore,%
%
%e6 ###
\begin{eqnarray}\label{eq:06}
\quad&&M_{1T}\bigl(P_{2}(t_{0})\leq\gamma\bigr)\nonumber\\
\quad&&\quad= M_{1T}\bigl(1-G_{1}\bigl((t_{0}-\mu_{0})^{2}/(1/n+\sigma_{2}^{2})\bigr)\leq\gamma\bigr)
\nonumber
\\[-8pt]
\\[-8pt]
\nonumber
\quad&&\quad= M_{1T}\biggl( \frac{(t_{0}-\mu_{0})^{2}}{1/n+\sigma_{1}^{2}}\geq\frac{%
    1/n+\sigma_{2}^{2}}{1/n+\sigma_{1}^{2}}G_{1}^{-1}(1-\gamma)\biggr)\\
\quad&&\quad= 1-G_{1}\biggl( \frac{1/n+\sigma_{2}^{2}}{1/n+\sigma_{1}^{2}}G_{1}^{-1}(1-\gamma)\biggr).\nonumber
\end{eqnarray}
We see immediately that \eqref{eq:06} will be less than $\gamma$ if
and only if $%
\sigma_{2}>\sigma_{1}$. In other words, $\Pi_{2}$ will be uniformly weakly
informative relative to $\Pi_{1}$ if and only if $\Pi_{2}$ is more diffuse
than $\Pi_{1}$. Note that $M_{1T}(P_{2}(t_{0})\leq\gamma)$ converges
to 0
as $\sigma_{2}^{2}\rightarrow\infty$ to reflect noninformativity. Also,
as $n\rightarrow\infty$, then \eqref{eq:06} increases to
$1-G_{1}((\sigma
_{2}^{2}/\sigma_{1}^{2})G_{1}^{-1}(1-\gamma))$. So we could ignore $n$
and choose $\sigma_{2}^{2}$ conservatively based on this limit, to obtain
an asymptotically uniformly weakly informative prior, as we know this value
of $\sigma_{2}^{2}$ will also be weakly informative for finite $n$.

If we specify that we want\vspace*{2pt} \eqref{eq:05} to equal $p\in\lbrack0,1]$,
then \eqref{eq:06}
implies that $\sigma_{2}^{2}=(1/n+\sigma_{1}^{2})(G_{1}^{-1}(1-\gamma
+p\gamma)/G_{1}^{-1}(1-\gamma))-1/n$. Such a choice will give a~proportion
$p$ fewer prior-data conflicts at level $\gamma$ than the base prior. This
decreases to $\sigma_{1}^{2}G_{1}^{-1}(1-\gamma+p\gamma
)/G_{1}^{-1}(1-\gamma)$ as $n\rightarrow\infty$ and so the more data we
have the less extra variance we need for $\Pi_{2}$ for weak informativity.

We can generalize this to $t\sim N_{k}(\mu,n^{-1}I)$ with $\Pi_{i}$ given
by $\mu\sim N_{k}(\mu_{0}, \Sigma_{i})$.
Note we have that $M_{iT}$ is the
$N_{k}(\mu_{0},n^{-1}I+\Sigma_{i})$ distribution. It is then easy to see
that $P_{2}(t_{0})=1-G_{k}((t_{0}-\mu_{0})^{\prime}(n^{-1}I+\Sigma
_{2})^{-1}(t_{0}-\mu_{0}))$ and
%
%e7 ###
\begin{eqnarray}\label{eq:07}
&&M_{1T}\bigl(P_{2}(t_{0})\leq\gamma\bigr)\nonumber\\
&&\quad=M_{1T}\bigl((t_{0}-\mu_{0})^{\prime
}(n^{-1}I+\Sigma_{2})^{-1}(t_{0}-\mu_{0})\\
&&\quad\hspace*{125pt}\geq G_{k}^{-1}(1-\gamma)\bigr).\nonumber
\end{eqnarray}
Note that \eqref{eq:07} increases to the probability that $(t_{0}-\mu
_{0})^{\prime
}\Sigma_{2}{}^{-1}(t_{0}-\mu_{0})\geq G_{k}^{-1}(1-\gamma)$, when $%
t_{0}\sim N_{k}(\mu_{0}, \Sigma_{1})$, as $n\rightarrow\infty$. This
probability can be easily computed via\vadjust{\goodbreak} simulation.

The following result is proved in the \hyperref[app]{Appendix}.\vspace*{-2pt}

\begin{theorem}\label{thm:01}
For a sample of $n$ from the statistical model
$\{N_{k}(\mu,I)\dvtx \mu\in R^{k}\}$, a $N_{k}(\mu_{0},\Sigma_{2})$
prior is
uniformly weakly informative relative to a $N_{k}(\mu_{0},\Sigma_{1})$
prior if and only if $\Sigma_{2}-\Sigma_{1}$ is positive
semidefinite.\vspace*{-2pt}
\end{theorem}

The necessary part of Theorem \ref{thm:01} is much more
difficult than the $k=1$ case and shows that we cannot have a $N_{k}(\mu_{0},\Sigma_{2})$
prior uniformly weakly informative relative to a $N_{k}(\mu_{0},\Sigma
_{1})$ prior unless \mbox{$\Sigma_{2}\geq\Sigma_{1}$}. It follows from
Theorem \ref{thm:01}
that a $N_{k}(\mu_{0},\Sigma_{2})$ prior is
uniformly weakly informative relative to a $N_{k}(\mu_{0},\Sigma_{1})$
prior if and only if a $N( a^t \mu_{0}, a^t\Sigma_{2}a)$ prior is
uniformly weakly informative relative to a $N(a^t\mu_{0},a^t\Sigma_{1}a)$
prior for every $a \in R^k$.

For the choice of $\Sigma
_{2}$ we have that, if $\Sigma_{1}$ and $\Sigma_{2}$ are arbitrary $%
k\times k$ positive definite matrices, then $r\Sigma_{2}\geq\Sigma_{1}$
whenever $r\geq\lambda_{k}(\Sigma_{1})/\lambda_{1}(\Sigma_{2})$
where $%
\lambda_{i}(\Sigma)$ deno\-tes the $i$th ordered eigenvalue of $\Sigma
$. Note that this condition does
not require that the $\Sigma_{i}$ have the same eigenvectors. When
they do have the
same eigenvectors, so $\Sigma_{i}=QD_{i}Q^{\prime}$ is the spectral
decomposition of $%
\Sigma_{i}$, then $\Sigma_{2}\geq\Sigma_{1}$ whenever $\lambda
_{i}(\Sigma_{2})\geq\lambda_{i}(\Sigma_{1})$ for $i=1,\ldots,k$.

%s3.2 ###
\vspace*{-2pt}\subsection{Comparing a $t$ Prior with a Normal Prior}\vspace*{-2pt}\label{sec:03.2}

It is not uncommon to find $t$ priors being substituted for normal
priors on
location parameters. Suppose $x=(x_{1},\ldots,x_{n})$ is a sample from
a $%
N(\mu,1)$ distribution where $\mu$ is unknown. We take $\Pi_{1}$ to
be a $%
N(\mu_{0},\sigma_{1}^{2})$ distribution and $\Pi_{2}$ to be a
$t_{1}(\mu
_{0},\sigma_{2}^{2},\lambda)$ distribution, that is, $t_{1}(\mu
_{0},\sigma
_{2}^{2},\lambda)$ denotes the distribution of $\mu_{0}+\sigma_{2}z$ with
$z$ distributed as a 1-di\-mensional $t$ distribution with $\lambda$ degrees
of freedom. We then want to determine $\sigma_{2}^{2}$ and $\lambda$
so that
the $t_{1}(\mu_{0},\sigma_{2}^{2},\lambda)$ prior is weakly informative
relative to the normal prior.

\begin{figure}

\includegraphics{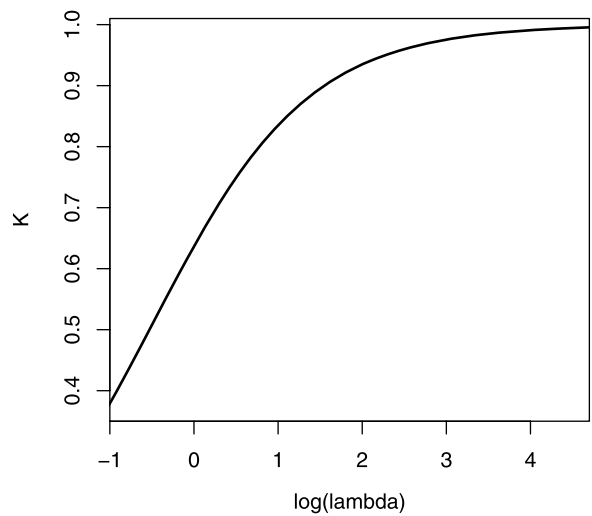}

  \caption{Plot of $ K(\lambda)$ against $\log(\lambda)$ where a $t_{1}(\mu
_{0},\sigma
_{2}^{2},\lambda)$ prior is asymptotically uniformly weakly
informative relative to a $N(\mu_{0},\sigma_{1}^{2})$ prior if and
only if $\sigma_{2}^{2}/\sigma_{1}^{2}\geq K(\lambda)$.}\label{fig1}
\end{figure}

We consider first the limiting case as $n\rightarrow\infty$. The limiting
prior predictive distribution of the minimal sufficient statistic
$T(x)=\bar{%
x}$ is $N(\mu_{0},\sigma_{1}^{2})$ while $P_{2}(t_{0})$ converges in
distribution to $1-H_{1,\lambda}((t_{0}-\mu_{0})^{2}/\sigma_{2}^{2})$
where $H_{1,\lambda}$ is the distribution function of an $F_{1,\lambda}$
distribution. This implies that \eqref{eq:04} converges to
$1-G_{1}((\sigma
_{2}^{2}/\sigma_{1}^{2})H_{1,\lambda}^{-1}(1-\gamma))$ and this is less
than or equal to $\gamma$ if and only if $\sigma_{2}^{2}/\sigma
_{1}^{2}\geq G_{1}^{-1}(1-\gamma)/H_{1,\lambda}^{-1}(1-\gamma))$. So to
have that $\Pi_{2}$ is asymptotically weakly informative relative to $%
\Pi_{1}$ at level $\gamma$, we must choose $\sigma_{2}^{2}$ large enough.
Clearly we have that $\Pi_{2}$ is asymptotically uniformly weakly
informative relative to $\Pi_{1}$ if and only if
\[
\sigma_{2}^{2}/\sigma_{1}^{2}\geq K(\lambda) = \sup_{\gamma\in
\lbrack
0,1]}G_{1}^{-1}(1-\gamma)/H_{1,\lambda}^{-1}(1-\gamma).
\]
In Figure \ref{fig1} we have plotted\vadjust{\goodbreak} $ K(\lambda)$ against $\log(\lambda)$.
%
%f1 ###

Since $K(1)=0.6366$, we require that $\sigma_{2}^{2}\geq\break \sigma
_{1}^{2}(0.6366)$ for a Cauchy prior to be uniformly weakly informative with
respect to a $N(\mu_{0},\sigma_{1}^{2})$ prior.\break A~$t_{1}(\mu
_{0},\sigma_{2}^{2},3)$ prior has variance $3\sigma_{2}^{2}$.
If we choose $\sigma_{2}^{2}$ so that the variance is $%
\sigma_{1}^{2}$, then $\sigma_{2}^{2}/\sigma_{1}^{2}=1/3$. Since
this is
less than $K(3)=0.8488$, this prior is not uniformly weakly
informative. A $%
t_{1}(\mu_{0},\sigma_{2}^{2},3)$ prior has to have variance at least equal
to $(2.5464)\sigma_{1}^{2}$ if we want it to be uniformly weakly
informative relative to a $N(\mu_{0},\sigma_{1}^{2})$ prior. This is
somewhat surprising and undoubtedly is caused by the peakedness of the~$t$
distribution. Note that $K(\lambda) \rightarrow1$ as $\lambda
\rightarrow
\infty$, so this increase in variance, for the $t$ prior over the normal
prior, decreases as we increase the degrees of freedom.

The situation for finite $n$ is covered by the following result proved
in the \hyperref[app]{Appendix}.%\smallskip

\begin{theorem}\label{thm:02}
For a sample of $n$ from the statistical model
$\{N(\mu,1)\dvtx \mu\in R^{1}\}$, a $t_{1}(\mu_{0},\sigma
_{2}^{2},\lambda)$
prior is uniformly weakly informative relative to a $N_{1}(\mu
_{0},\sigma_{1}^{2})$ prior whenever $\sigma_{2}^{2}\geq\sigma_{0n}^{2}$,
where $\sigma_{0n}^{2}$ is the unique solution of $(1/n+\sigma
_{1}^{2})^{-1/2}=\int_{0}^{\infty}(1/n+\sigma
_{0n}^{2}/u)^{-1/2}\cdot k_{\lambda
}(u) \,du$ with $k_{\lambda}$ the $\operatorname{Gamma}_{\mathrm
{rate}}(\lambda
/2,\lambda
/2) $ density. Further, $\sigma_{0n}^{2}/\sigma_{1}^{2}$ increases to
%
%e8 ###
\begin{equation}
\quad K(\lambda) = \sup_{\gamma\in\lbrack0,1]}\frac{G_{1}^{-1}(1-\gamma
)}{H_{1,\lambda
}^{-1}(1-\gamma)}=\frac{2}{\lambda}\frac{\Gamma^{2}((\lambda+1)/2)}{
\Gamma^{2}(\lambda/2)} \label{eq:08}
\end{equation}
as $n\rightarrow\infty$ and so a $t_{1}(\mu_{0},\sigma
_{2}^{2},\lambda)$
prior is asymptotically uniformly weakly informative if and only if
$\sigma
_{2}^{2}/\sigma_{1}^{2}$ is greater than or equal to \eqref{eq:08}.%
\end{theorem}

Theorem \ref{thm:02} establishes that we can conservatively
use \eqref{eq:08} to select
a uniformly weakly informative $t$ prior.

\begin{figure}

\includegraphics{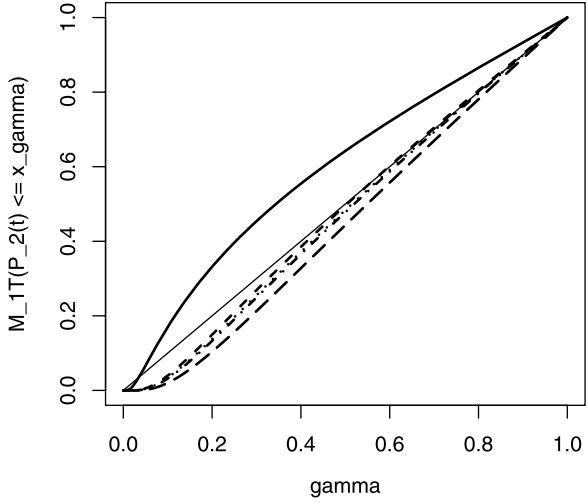}

  \caption{Plot of (\protect\ref{eq:04}) versus $\protect\gamma$ for $t_{1}(0,\sigma_{2}^{2},3)$
priors relative to a $N(0,1)$ prior when $n=20$,
where $\protect\sigma_{2}^{2}$ is chosen to match variances (thick
solid line),
match the MAD (dashed line), just achieve uniform weak
informativity (dotted line), just achieve
asymptotic uniform weak informativity (dash-dot line), and equal to 1
(long\mbox{-}dashed line).}\label{fig2}
\end{figure}

In Figure \ref{fig2} we have plotted the value of \eqref{eq:04} that arises with
$%
t_{1}(0,\sigma_{2}^{2},3)$ priors, where $\sigma_{2}^{2}$ is chosen
in a
variety of ways, together with the 45-degree line. A uniformly weakly
informative prior will have \eqref{eq:04} always below the 45-degree
line, while a
uniformly weakly informative prior at level $\gamma_{0}$ will have
\eqref{eq:04}
below the 45-degree line to the left of $\gamma_{0}$ and possibly
above to
the right of $\gamma_{0}$. For example, when $\sigma_{2}^{2}=1/3$,
then the $t_{1}(0,\sigma_{2}^{2},3)$ prior and the $N(0,1)$ prior have the
same variance. We see that this prior is only uniformly weakly informative
at level $\gamma_{0}=0.0357$ and is not uniformly weakly informative.

%f2 ###

Note that \eqref{eq:05} converges to $1-G_{1}((\sigma_{2}^{2}/\sigma
_{1}^{2})H_{1,\lambda}^{-1}(1-\gamma))/\gamma$ as $n\rightarrow
\infty$,
and setting this equal to $p$ implies that $\sigma_{2}^{2}=\sigma
_{1}^{2}G_{1}^{-1}(1-\gamma+\gamma p)/H_{1,\lambda}^{-1}(1-\gamma)$ which
converges, as $\lambda\rightarrow\infty$, to the result we obtained in
Section~\ref{sec:03.1}. So when $\lambda=3,\gamma=0.05$ and $p=0.5$,
we must have $%
\sigma_{2}^{2}/\sigma_{1}^{2}=5.0239/10.1280=0.49604$.

Our analysis indicates that one has to be careful about the scaling of
the $%
t $ prior if we want to say that the $t$ prior is less informative than a
normal prior, at least when we want uniform weak informativity.

Consider now comparing a multivariate $t$ prior to a multivariate normal
prior. Let $t_{k}(\mu_{0},\Sigma_{2},\lambda)$ denote the $k$-dimensional
$t$ distribution given by $\mu_{0}+\Sigma_{2}^{1/2}z$, where $\Sigma
_{2}^{1/2}$ is a square root of the positive definite matrix $\Sigma_{2}$
and $z$ has a $k$-dimensional $t$ distribution with $\lambda$ degrees of
freedom. This is somewhat more complicated than the normal case, but we
prove the following result in the \hyperref[app]{Appendix} which provides sufficient
conditions for the asymptotic uniform weak informativity.

\begin{theorem}\label{thm:03}
When sampling from the statistical model $\{N_{k}(\mu,I)\dvtx \mu\in R^{k}\}$,
a $t_{k}(\mu_{0},\Sigma_{2},\lambda)$
prior is asymptotically uniformly weakly informative relative to a $%
N_{k}(\mu_{0},\Sigma_{1})$ prior whenever\ $\Sigma_{2}-\tau_{\lambda
}^{2}\Sigma_{1}$ is positive semidefinite, where $\tau_{\lambda
}^{2}=(2/\lambda)\Gamma^{2/k}((k+\lambda)/2)/\allowbreak\Gamma^{2/k}(\lambda/2)$.
\end{theorem}

In contrast with Theorem \ref{thm:01}, we do not have an
equivalent characterization
of the uniform weak informativity of multivariate $t$ priors in terms of
the marginal priors of $a^t\mu$. For example, when $%
k=2$, then $\tau_{\lambda}^{2}\,{=}\,1$ and when $k\,{=}\,1$, then $\tau_{\lambda
}^{2}\,{=}\,2\Gamma^{2}((\lambda\,{+}\,1)/2)/\break\lambda\Gamma^{2}(\lambda/2)<1$ for
all $\lambda$. Therefore, $a^{t}\Sigma_{2}a-\{2\Gamma^{2}((\lambda
+1)/2)/\lambda\Gamma^{2}(\lambda/2)\}a^{t}\Sigma_{1}a>0$ for all
$a$ does
not imply that $\Sigma_{2}-\Sigma_{1}$ is positive semidefinite, for example,
take $\Sigma_{2}=\Sigma_{1}(1+2\Gamma^{2}((\lambda+1)/2)/ \lambda
\Gamma
^{2}(\lambda/2))/ 2$.

For the choice of $\Sigma_{2}$ we have that, if $\Sigma_{1}$ and
$\Sigma_{2}$ are arbitrary $k\times k$ positive definite matrices,
then $%
r\Sigma_{2}\geq\tau_{\lambda}^{2}\Sigma_{1}$ whenever $r\geq\tau
_{\lambda}^{2}\lambda_{k}(\Sigma_{1})/\lambda_{1}(\Sigma_{2})$. When
the $\Sigma_{i}$ have the same eigenvectors,
then $\Sigma_{2}\geq\tau_{\lambda}^{2} \Sigma_{1}$ whenever $\lambda
_{i}(\Sigma_{2})\geq\tau_{\lambda}^{2}\lambda_{i}(\Sigma_{1})$
for $%
i=1,\ldots,k$.

%s3.3 ###
\vspace*{2pt}\subsection{Comparing Inverse Gamma Priors}\vspace*{2pt}\label{sec:03.3}

Suppose now that we have a sample $x\,{=}\,(x_{1},\ldots,x_{n})$ from a $%
N(0,\sigma^{2})$ distribution where $\sigma^{2}$ is unknown. Then $%
t=T(x)=(x_{1}^{2}+\cdots+x_{n}^{2})/n$ is minimal suffi\-cient and
$T\sim\operatorname{Gamma}_{\mathrm{rate}}(n/2,n/2\sigma^{2})$.
Now suppose that we take $\Pi
_{i} $ to be an inverse gamma prior on $\sigma^{2}$, namely, $\sigma
^{-2}\sim\operatorname{Gamma}_{\mathrm{rate}}(\alpha_{i},\beta_{i})$.
From this we get
that $\alpha_{i}T/\beta_{i}\sim F(n,2\alpha_{i})$ and, since $%
J_{T}(x)=(4x^{\prime}x/\break n)^{-1/2}=(4t/n)^{-1/2},m_{iT,n}^{\ast
}(t)=m_{iT,n}(t) \cdot (4t/n)^{1/2}\propto t^{(n-1)/2}(1+nt/2\beta
_{i})^{-n/2-\alpha_{i}}$, which implies
\begin{eqnarray*}
P_{i,n}(t_{0}) &=& M_{iT,n}\bigl(t^{(n-1)/2}(1 + nt/2\beta_{i})^{-n/2-\alpha
_{i}}\\
&&\hspace*{30pt}\leq
t_{0}^{(n-1)/2}(1 + nt_{0}/2\beta_{i})^{-n/2-\alpha_{i}}\bigr).
\end{eqnarray*}

%f3 ###
\begin{figure*}

\includegraphics{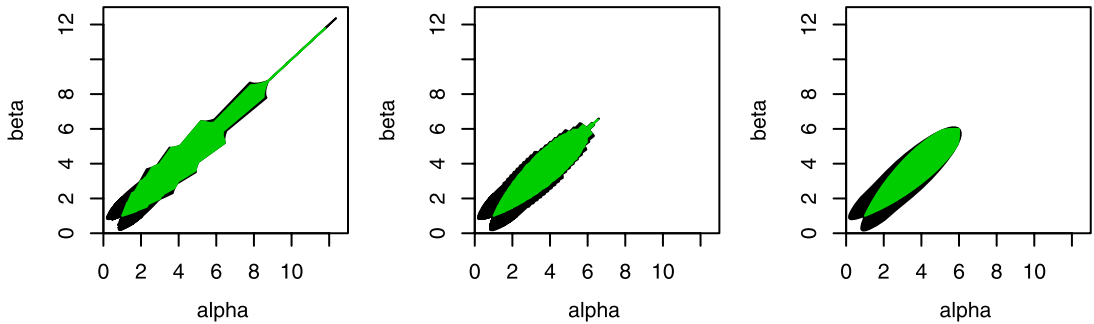}

  \caption{Plot of $(\alpha,\beta)$ corresponding to
$\operatorname{Beta}(\alpha,\beta)$ priors
that are weakly informative
at level $\gamma=0.05$ (light and dark shading)
and uniformly weakly informative (light shading) for $n=20$ (on the left), $n=100$
(middle) and $n=\infty$ (on the right).}\label{fig3}
\end{figure*}

We want to investigate the weak informativity of a~$\operatorname
{Gamma}_{\mathrm{rate}%
}(\alpha_{2},\beta_{2})$ prior relative to a $\operatorname
{Gamma}_{\mathrm{rate}}(\alpha_{1},\break \beta_{1})$ prior. For finite $n$ this is a difficult problem, so
we simplify this by considering only the asymptotic case. When the
prior is $%
\Pi_{i}$, then, as $n\rightarrow\infty$, we have that $%
m_{iT,n}(t)\rightarrow m_{iT}(t)=(\beta_{i}^{\alpha_{i}}/\Gamma
(\alpha
_{i}))t^{-\alpha_{i}-1}e^{-\beta_{i}/t}$, that is, $1/t\sim\operatorname
{Gamma}_{\mathrm{%
rate}}(\alpha_{i},\beta_{i})$ in the limit.\break \mbox{Therefore}, $%
P_{2,n}(t_{0})\rightarrow P_{2}(t_{0})=\Pi_{2}(t^{-\alpha
_{2}-1/2} e^{-\beta_{2}/t}\leq\break t_{0}^{-\alpha_{2}-1/2}\cdot e^{-\beta
_{2}/t_{0}})$ and we want to determine
conditions on $(\alpha_{2},\beta
_{2})$ so that $\Pi_{1}(P_{2}(t)\leq\gamma)\leq\gamma$.

While results can be obtained for this problem, it is still rather difficult.
It is greatly simplified,\vadjust{\goodbreak} however, if we impose a natural restriction
on $%
(\alpha_{2},\beta_{2})$. In particular, we want the location of the bulk
of the mass for $\Pi_{2}$ to be located roughly in the same place as the
bulk of the mass for $\Pi_{1}$. Accordingly, we could require the
priors to
have the same means or modes, but, as it turns out, the constraint that
requires the modes of the $m_{iT}^{\ast}$ functions to be the same greatly
simplifies the analysis. Actually,\vspace*{2pt} $m_{iT,n}^{\ast}(t)$ converges to
0, but
the $n$'s cancel in the inequalities defining~$P_{i,n}(t_{0})$ and so
we can
define $m_{iT,n}^{\ast}(t)=\break t^{-\alpha_{i}-1/2}e^{-\beta_{i}/t}$
which has
its mode at $t=\beta_{i}/(\alpha_{i}+1/2)$. Therefore, we must have
$\beta
_{2}/(\alpha_{2}+1/2)=\beta_{1}/\break(\alpha_{1}+1/2)$ so that $(\alpha
_{2},\beta_{2})$ lies on the line through the points $(0,\beta
_{1}/2(\alpha_{1}+1/2))$ and $(\alpha_{1},\beta_{1})$. We prove the
following result in the \hyperref[app]{Appendix}.%\smallskip

\begin{theorem}\label{thm:04}
Suppose we use a $\operatorname{Gamma}_{\mathrm{rate}}(\alpha_{1},\break\beta_{1})$
prior on $1/\sigma^{2}$ when sampling from the
statistical model $\{N(0,\sigma^{2})\dvtx \sigma^{2}>0\}$. Then a
$\operatorname{Gamma}_{\mathrm{rate}}(\alpha_{2}, \beta_{2})$
prior on $1/\sigma^{2}$, with $\beta_{2}/(\alpha_{2}+1/2)=\beta_{1}/(\alpha_{1}+1/2)$, is
asymptotically weakly informative relative to the $\operatorname
{Gamma}_{\mathrm{rate}}(\alpha_{1},\beta_{1})$ prior whenever $\alpha_{2}\leq\alpha_{1}$
and $\beta_{2}=\beta_{1}(\alpha_{2}+1/2)/(\alpha_{1}+1/2)$ or,
equivalently, whenever $\beta_{1}/2(\alpha_{1}+1/2)\leq\beta_{2}\leq
\beta_{1}$ and $\alpha_{2}=(\alpha_{1}+1/2)\beta_{2}/ \beta
_{1}-1/2$.% \smallskip
\end{theorem}

Of particular interest here is that we cannot reduce the rate
parameter $\beta_{2}$ arbitrarily close to 0 and be guaranteed asymptotic
weak informativity.

%s4 ###
\section{Applications}\label{sec:04}

We consider now some applications of determining weakly informative priors.

%s4.1 ###
\subsection{Weakly Informative Beta Priors for the Binomial}\label{sec:04.1}

Suppose that $T\!\sim\!\operatorname{Binomial}(n,\theta)$ and $\theta\!\sim\! \operatorname{Beta}(\alpha,\beta)$.
This implies that $m_{T}(t)=\bigl({{n }\atop{t}}\bigr)\Gamma(\alpha+\beta)\Gamma(t+\alpha)\Gamma(n-t+\beta)/
\Gamma(\alpha)\Gamma(\beta)\Gamma(n+\alpha+\beta)$ and from this we can compute \eqref{eq:04} for various choices
of $(\alpha,\beta)$.

As a specific example, suppose that $n=20$, the base prior is given by $
(\alpha,\beta)=(6,6)$, and we take $\gamma=0.05$ so that $x_{0.05}=0.0588$.
As alternatives to this base prior, we consider $\operatorname{Beta}(\alpha,\beta)$
priors. In Figure \ref{fig3} we have plotted all the $(\alpha,\beta)$ corresponding
to $\operatorname{Beta}(\alpha,\beta)$ distributions that are weakly informative with
respect to the $\operatorname{Beta}(6,6)$ distribution at level $0.05$, together with the
subset of all $(\alpha,\beta)$ corresponding to $\operatorname{Beta}(\alpha,\beta)$
distributions that are uniformly weakly informative relative to the
$\operatorname{Beta}
(6,6)$ distribution. The graph on the left corresponds to $n=20$, the middle
graph corresponds to $n=100$, and the graph on the right corresponds to
$%
n=\infty$. The plot for $n=20$ shows some anomalous effects due to the
discreteness of the prior predictive distributions and these effects
disappear as $n$ increases.\ In such an application we may choose to
restrict to symmetric priors, as this fixes the primary location of the prior
mass. For example, when $n=20$, a $\operatorname{Beta}(\alpha,\alpha)$ prior for
$\alpha$
satisfying $1\leq\alpha\leq12.3639$ is uniformly weakly informative with
respect to the $\operatorname{Beta}(6,6)$ prior and we see that values of $\alpha> 6$ are
eliminated as $n$ increases.

%s4.2 ###
\subsection{Weakly Informative Priors for the Normal Regression Model}\label{sec:04.2}

Consider the situation where $y \sim N_n(X \beta,\sigma^2 I)$, $X \in
R^{n \times k}$ is
of rank $k$ and $\beta\in R^k, \sigma^2>0$ are unknown. Therefore,
$T=(b,s^2)$ with $b=(X'X)^{-1}X'y$ and $s^2=\Vert y-Xb\Vert ^2$.
Suppose we have elicited a prior on $(\beta,\sigma^{2})$ given by\vadjust{\goodbreak}
$1/\sigma^{2}\sim\operatorname{Gamma}_{\mathrm{rate}}(\alpha_{1},\tau
_{1})$, and
$\beta| \sigma^{2}\sim N_k(\beta_{0},\sigma^{2} \Sigma_1)$.
We now find a prior that is asymptotically uniformly weakly
informative relative to this choice. For this we consider gamma priors
for $1/\sigma^{2}$ and $t$ priors for $\beta$ given $\sigma^{2}$.
For the asymptotics we suppose that $\lambda_k ((X'X)^{-1}) \rightarrow
0$ as
$n \rightarrow\infty$.

As discussed in Evans and Moshonov
(\citeyear{EvansMoshonov2006,EvansMoshonov2007}), it seems that
the most sensible way to check for prior-data
conflict here is to first check the prior on~$\sigma^{2}$, based on the
prior predictive distribution of~$s^{2}$. If no prior-data conflict is
found at this stage, then we check the prior on $\beta$ based on the conditional
prior predictive for $b$ given $s^{2}$, as $s^{2}$ is ancillary for
$\beta$.
Such an approach provides more information concerning where a prior-data
conflict exists than simply checking the whole prior via \eqref{eq:03}.

So we consider first obtaining an asymptotically uniformly weakly
informative prior for $1/\sigma^{2}$.
We have that $s^{2} | \sigma^{2}\sim\operatorname{Gamma}_{\mathrm{rate}%
}((n-k)/2,(n-k)/2\sigma^{2})$ and so, as in Section \ref{sec:03.3}, when
$1/\sigma^{2}\sim\operatorname{Gamma}_{\mathrm{rate}}(\alpha_{i},\break\tau_{i})$, the
limiting prior
predictive distribution of $1/s^{2}$ is $\operatorname{Gamma}_{\mathrm
{rate}}(\alpha
_{i},\tau_{i})$ as $n\rightarrow\infty$. Furthermore, when $T_2(x)=s^{2}$,
then $J_{T_2}(x)=(4s^{2}/(n-k))^{-1/2}$. Therefore, the limiting value
of \eqref{eq:04}
in this case is the same as that discussed in Section \ref{sec:03.3}
and Theorem \ref{thm:04}
applies to obtain a $\operatorname{Gamma}_{\mathrm{rate}}(\alpha
_{2},\tau_{2})$ prior
asymptotically
uniformly weakly informative relative to the $\operatorname
{Gamma}_{\mathrm{rate}}(\alpha
_{1},\tau_{1})$ prior.

If we consider $s^{2}$ as an arbitrary fixed value from its prior predictive
distribution, then, when
$\beta| \sigma^{2}\sim N_k(\beta_{0},\sigma^{2} \Sigma_1 )$,
the conditional prior predictive distribution of
$b$ given $s^{2}$ converges to the $N_k(\beta_{0},s^{2} \Sigma_1)$
distribution. Furthermore, when
$\beta| \sigma^{2}\sim t_k(\beta_{0},\sigma^{2} \Sigma_2,\break \lambda)$,
the conditional prior predictive distribution of
$b$ given~$s^{2}$ converges to the $t_k(\beta_{0},s^{2} \Sigma_2,\lambda)$
distribution. So we can apply Lemma \ref{lem:01} to these limiting
distributions. It is then clear that
the comparison is covered by Theorem \ref{thm:03}, as the limiting
prior predictives are of the same form.
Therefore, the $t_k(\beta_{0},\sigma^{2} \Sigma_2, \lambda)$
prior is asymptotically uniformly weakly informative
relative to the $N_k(\beta_{0},\sigma^{2} \Sigma_1)$ prior
whenever $s^{2} \Sigma_2 \geq s^{2} \tau_{\lambda
}^{2}\Sigma_1 $ or, equivalently, whenever
$\Sigma_2\!\geq\!\tau_{\lambda}^{2}\Sigma_1 $ where~$\tau_{\lambda
}^{2}$ is defined in Theorem \ref{thm:03}.
Note that this condition does not depend on $s^{2}$.
Also, as $\lambda\rightarrow\infty$, we can use Theorem \ref{thm:02}
to obtain that a
$N_k(\beta_{0},\sigma^{2} \Sigma_2)$ prior is asymptotically
uniformly weakly informative
relative to the $N_k(\beta_{0},\sigma^{2} \Sigma_1)$ prior
whenever
$\Sigma_2 \geq\Sigma_1 $.

%s4.3 ###
\subsection{Weakly Informative Priors for Logistic Regression}\label{sec:04.3}

Supposing we have a single binary valued response variable\vadjust{\goodbreak} $Y$ and $k$
quantitative predictors $X_{1},\ldots,X_{k}$, we observe
$(Y,X_{1},\ldots,X_{k})$ at $q$ settings of the predictor variables and have
$n_{i}$ observations at the $i$th setting of the predictors. The logistic
regression mo\-del then says that $Y_{ij}\sim\operatorname{Bernoulli}(p_{i})$
where $\log(p_{i}/\break(1-p_{i}))=\beta_{0}+\beta_{1}(x_{i1}-\bar{x}_{\cdot1
})+\cdots+\beta_{k}(x_{ik}-\bar{x}_{\cdot k})$ for $j=1,\ldots
,n_{i}$ and
$i=1,\ldots,q$ and the $\beta_{i}$ are unknown real values. For simplicity,
we will assume no $x_{i j} - \bar{x}_{\cdot j}$ is zero. For this model
$%
T=(T_{1},\ldots,T_{q})$, with $T_{i}=Y_{i1}+\cdots+Y_{in_{i}}$, is a
minimal sufficient statistic. For the base prior we suppose that $\Pi_{1}$
is the product of independent priors on the $\beta_{i}$'s and we consider
the problem of finding a prior $\Pi_{2}$ that is weakly informative
relative to $\Pi_{1}$. For example, we could take~$\Pi_{1}$ to be a
product of $N(0,\sigma_{1i}^{2})$ priors and~$\Pi_{2}$ to be a
product of $%
N(0,\sigma_{2i}^{2})$ priors and choose the~$\sigma_{2i}^{2}$ so that weak
informativity is obtained. Note that since~$T$ is discrete we can use
\eqref{eq:02} in
our computations.

As we will see, it is not the case that choosing the $\sigma_{2i}^{2}$ very
large relative to the $\sigma_{1i}^{2}$ will necessarily make $\Pi_{2}$
weakly informative relative to $\Pi_{1}$. In fact, there is only a
finite range of $\sigma_{2i}^{2}$ values where weak informativity will
obtain.

While this can be demonstrated analytically, the argument is somewhat
technical and it is perhaps easier to see this in an example. The following
bioassay data are from Racine et al. (\citeyear{Racine-etal1986})
and were also analyzed in Gelman et al. (\citeyear{Gelman-etal2008}).
These data arise from an experiment where 20 animals were
exposed to four doses of a toxin and the number of deaths recorded (Table \ref{tab1}).

%t1 ###
\begin{table}
\caption{}\label{tab1}
\begin{tabular*}{\columnwidth}{@{\extracolsep{\fill}}lcc@{}}
\hline
\multicolumn{1}{@{}l}{\textbf{Dose (g}$\bolds{/}$\textbf{ml)}} & \multicolumn{1}{c}{\textbf{Number of animals} $\bolds{n_{i}}$} &
\multicolumn{1}{c@{}}{\textbf{Number of deaths} $\bolds{t_{i}}$} \\
\hline
{0.422} & {5} & {0}
\\
{0.744} & {5} & {1}
\\
{0.948} & {5} & {3}
\\
{2.069} & {5} & {5}
\\
\hline
\end{tabular*}
\end{table}
%
%}%
%%{Table 1. \it Bioassay data from Racine et al. (1986).}
%

Following Gelman et al. (\citeyear{Gelman-etal2008}), we
took $X_{1}$ to be the variable formed by
calculating the logarithm of dose and then standardizing to make the
mean of
$X_{1}$ equal to 0 and its standard deviation equal to 1$/$2. Gelman
et al.
(\citeyear{Gelman-etal2008}) placed independent Cauchy priors on the
regression
coefficients, namely, $\beta_{0}\sim t_{1}(0,10^{2},1)$ independent of
$\beta
_{1}\sim t_{1}(0,2.5^{2},1)$.

\begin{figure*}
\centering
\begin{tabular}{@{}cc@{}}

\includegraphics{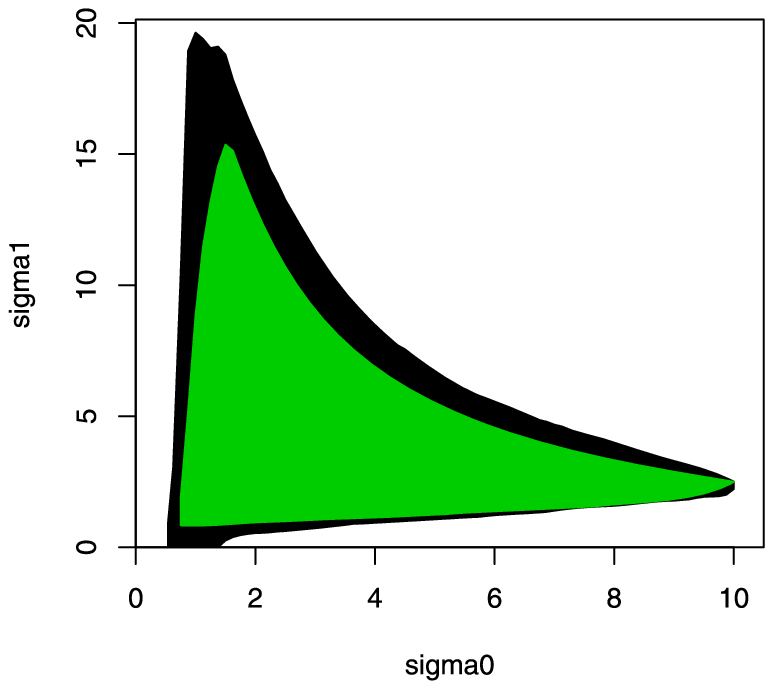}
 & \includegraphics{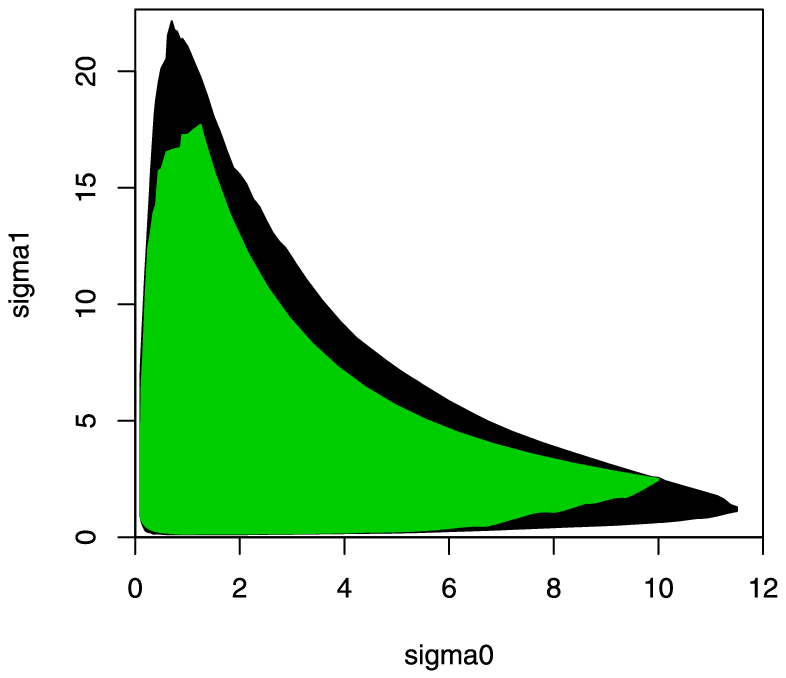}\\
\footnotesize{(a)} & \footnotesize{(b)}\\

\includegraphics{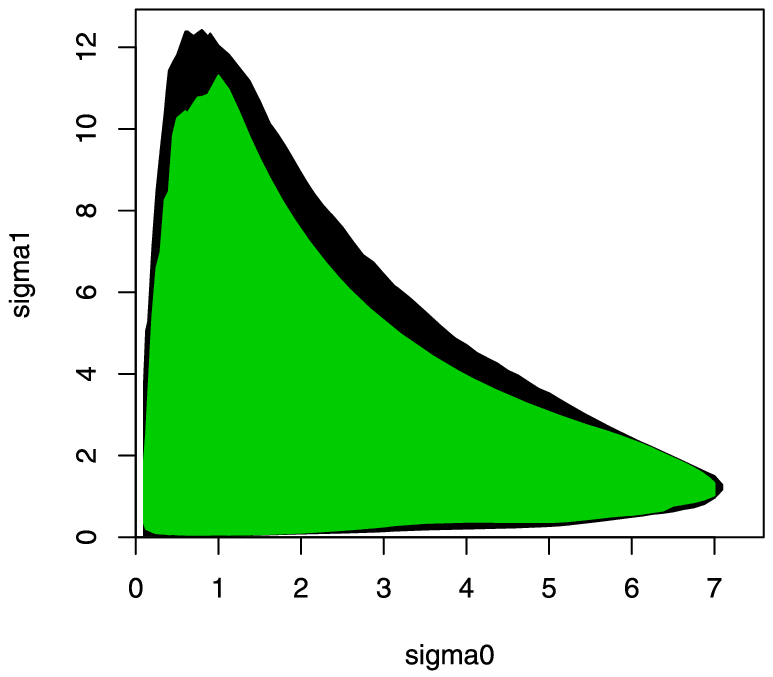}
 & \includegraphics{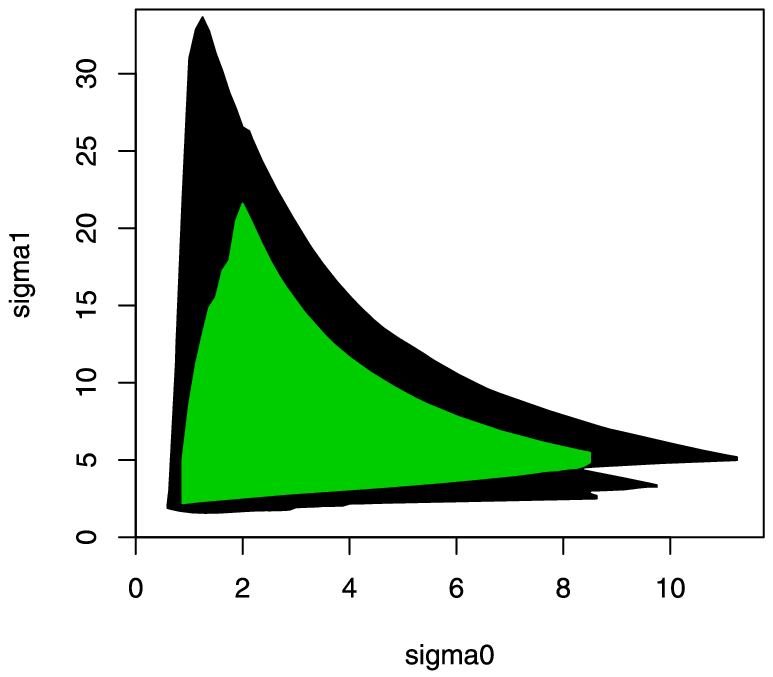}\\
\footnotesize{(c)} & \footnotesize{(d)}
\end{tabular}
 \vspace*{-3pt}
 \caption{Weakly informative $%
\Pi_{2}$ priors relative to $\Pi_{1}$ at level 0.05 (light and dark
shading) and uniformly weakly informative (light shading) where
(\textup{a}) $\Pi_{1}=N(0,10^{2})\times N(0,2.5^{2})$ and $\Pi_{2}=N(0,\sigma
_{0}^{2})\times N(0,\sigma_{1}^{2})$,
(\textup{b}) $\Pi_{1}= t_{1}(0,10^{2},1)\times t_{1}(0,2.5^{2},1)$ and $\Pi
_{2}=t_{1}(0,\sigma_{0}^{2},1)\times t_{1}(0,\sigma_{1}^{2},1)$,
(\textup{c})~$\Pi_{1}=N(0,10^{2})\times N(0,2.5^{2})$ and $\Pi_{2}=t_{1}(0,\sigma
_{0}^{2},1)\times t_{1}(0,\sigma_{1}^{2},1)$ and
(\textup{d})~$\Pi_{1}=t_{1}(0,10^{2},1)\times\allowbreak  t_{1}(0,2.5^{2},1)$ and $\Pi
_{2}= N(0,\sigma_{0}^{2})\times N(0,\sigma_{1}^{2})$.}\label{fig4}
\vspace*{-6pt}
\end{figure*}

We consider four possible scenarios for the investigation of weak
informativity at level $\gamma=0.05$ and uniform weak informativity. In
Figure \ref{fig4}(a) we compare $\Pi_{2}=N(0,\sigma_{0}^{2})\times N(0,\sigma
_{1}^{2})$ priors with the prior $\Pi_{1}=N(0,10^{2})\times N(0,2.5^{2})$.
The entire region gives the $(\sigma_{0},\sigma_{1})$ values corresponding
to priors\vadjust{\goodbreak} that are weakly informative at level $\gamma=0.05$, while the
lighter subregion gives the $(\sigma_{0},\sigma_{1})$ values corresponding
to priors that are uniformly weakly informative. Note that some of the
irregularity in the plots is caused by the fact that the prior predictive
distributions of $T$ are discrete. The three remaining plots are similar
where in Figure \ref{fig4}(b) $\Pi_{1}=t_{1}(0,10^{2},1)\times
t_{1}(0,2.5^{2},1)$ and $\Pi
_{2}=t_{1}(0,\sigma_{0}^{2},1)\times t_{1}(0,\sigma_{1}^{2},1)$, in
Figure \ref{fig4}(c) $%
\Pi_{1}=N(0,10^{2})\times N(0,2.5^{2})$ and $\Pi_{2}=t_{1}(0,\break\sigma
_{0}^{2},1)\times t_{1}(0,\sigma_{1}^{2},1)$, and in Figure \ref{fig4}(d) $\Pi
_{1}=t_{1}(0,10^{2},\break 1)\times t_{1}(0,2.5^{2},1)$ and $\Pi
_{2}=N(0,\sigma
_{0}^{2})\times N(0,\sigma_{1}^{2})$. Note that these plots only
depend on
the data through the values\vadjust{\goodbreak} of $X_{1}$.
%
%f4 ###

%
%%
%&\includegraphics[scale=.4]{wi-fig-logis-cc.eps} \\
%(a) & (b) \\[2ex]
%&\includegraphics[scale=.4]{wi-fig-logis-cn.eps} \\
%(c) & (d)
%%
%}
% }
%%original-width 6.3183in;original-height 6.2059in;

We see clearly from these plots that increasing the scaling on any of
the $%
\beta_{i}$ does not necessarily\vadjust{\goodbreak} lead to weak informativity and in fact
inevitably destroys it. Furthermore, a smaller scaling on a parameter can
lead to uniform weak informativity. These plots underscore how our intuition
does not work very well with the logistic regression model, as it is not
clear how priors on the $\beta_{i}$ ultimately translate to priors on
the $p_{i}$.
In fact, it can be proven that, if we put independent priors on the
$\beta_{i}$,
fix all the scalings but one, and let that scaling grow arbitrarily large,
then the prior predictive distribution of $T$ converges to a~distribution
concentrated on two points, for example, when the scaling
on $\beta_0$ increases these points are given by $\{\sum
_{i=1}^{q}T_{i}=0\}\cup
\{\sum_{i=1}^{q}T_{i}=\sum_{i=1}^{q}n_{i}\}$, and this is definitely not
desirable. This partially explains the results obtained.

\begin{figure}

\includegraphics{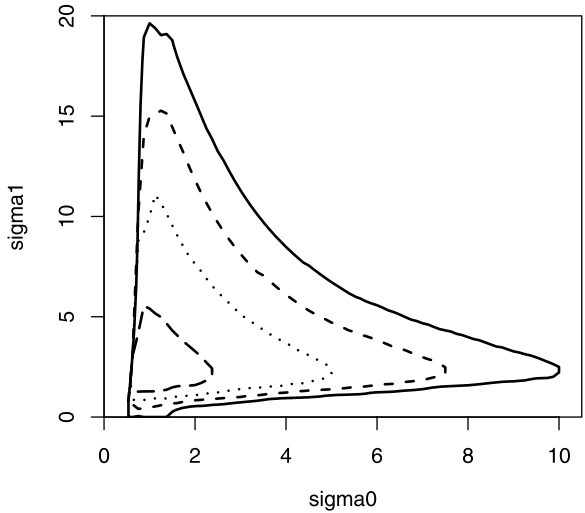}

  \caption{Reduction levels of $N(0,\protect\sigma_{0}^{2})\times N(0,\protect
\sigma_{1}^{2})$
relative to $N(0,10^{2})\times N(0,2.5^{2})$ priors using (\protect\ref{eq:05})
when $\protect\gamma=0.05$.
The plotted reduction levels are 0\% (solid line), 25\%
(dashed line), 50\% (dotted line) and 75\% (long dashed line).}\label{fig5}
\end{figure}

Of some interest is how much reduction we actually get, via \eqref
{eq:05}, when we
employ a weakly informative prior. In Figure \ref{fig5} we have plotted contours of
the choices of $(\sigma_{0},\sigma_{1})$ that give 0\%, 25\%, 50\% and
75\% reduction in prior-data conflicts for the case where $\Pi
_{2}=N(0,\sigma_{0}^{2})\times N(0,\sigma_{1}^{2})$ and $\Pi
_{1}=N(0,10^{2})\times N(0,2.5^{2})$ when $\gamma=0.05$ (this corresponds
to $x_{\gamma}=0.0503$). Note that a substantial reduction can be
obtained.%

%f5 ###

We can also consider
fixing one of the scalings and seeing how much reduction we obtain when
varying the other.\ For example, when we fix $\sigma_{0}=2.5$ we find that
the maximum reduction is obtained when~$\sigma_{1}$ is close to 2.2628,
while if we fix $\sigma_{1}=2.5$, then the maximum reduction is obtained
when $\sigma_{0}$ is close to~0.875.

It makes sense in any application to check to see if any prior-data conflict
exists with respect to the base prior. If there is no prior-data conflict,
this increases our confidence that the weakly informative prior is indeed
putting less information into the analysis. This is assessed generally using
\eqref{eq:03}, although \eqref{eq:02} suffices in this example. When
$\Pi_{1}=N(0,10^{2})\times
N(0,2.5^{2})$, then \eqref{eq:02} equals $0.1073$\ and when $\Pi
_{1}=t_{1}(0,10^{2},1)\times t_{1}(0,2.5^{2},1)$ (the prior used in
Gelman et al., \citeyear{Gelman-etal2008}),\vadjust{\goodbreak}
then \eqref{eq:02} equals $0.1130$, so in neither case is there any evidence
of prior-data conflict.

%s5 ###
\section{Refinements Based Upon Ancillarity}\label{sec:05}

Consider an ancillary statistic that is a function of the minimal
sufficient statistic,
say, $U(T)$. The variation due to $U(T)$ is independent of $\theta$
and so
should be removed from the $P$-value \eqref{eq:03} when checking for prior-data
conflict. Removing this variation is equivalent to conditioning on $U(T)$
and so we replace \eqref{eq:03} by%
%
%e9 ###
\begin{equation}
M_{T}\bigl(m_{T}^{\ast}(t)\leq m_{T}^{\ast}(t_{0}) | U(T)\bigr), \label{eq:09}
\end{equation}
that is, we use the conditional prior predictive given the ancillary
$U(T)$. To
remove the maximal amount of ancillary variation, we must have that
$U(T)$ is a~maximal ancillary. Therefore, \eqref{eq:04} becomes
%
%e10 ###
\begin{equation}
M_{1T}\bigl(P_{2}(t_{0} | U(T))\leq x_{\gamma} | U(T)\bigr), \label{eq:10}
\end{equation}
that is, we have replaced $P_{2}(t_{0})$ by $P_{2}(t_{0} |
U(T))=M_{2T}(m_{2T}^{\ast}(t)\leq m_{2T}^{\ast}(t_{0}) | U(T))$
and $M_{1T}$ by\break $M_{1T}(\cdot|  U(T))$.

We note that the approach discussed in Section \ref{sec:02}
works whenever $T$ is a complete minimal
sufficient statistic. This is a consequence of Basu's Theorem, as, in
such a
case, any ancillary is statistically independent of $T$ and so conditioning
on such an ancillary is irrelevant. This is the case for the examples
in Sections \ref{sec:03}~and~\ref{sec:04}.

One problem with ancillaries is that multiple maximal ancillaries may exist.
When ancillaries are used for frequentist inferences about $\theta$ via
conditioning, this poses a problem because it is not clear which maximal
ancillary to use and confidence regions depend on the maximal ancillary
chosen. For checking for prior-data conflict via \eqref{eq:09},
however, this does not
pose a problem. This is because we simply get different checks
depending on
which maximal ancillary we condition on. For example, if conditioning on
maximal ancillary $U_{1}(T)$ does not lead to prior-data conflict, but
conditioning on maximal ancillary $U_{2}(T)$ does, then we have evidence
against no prior-data conflict existing.

Similarly, when we go to use \eqref{eq:10}, we can also simply look at
the effect of
each maximal ancillary on the analysis and make our assessment about
$\Pi
_{2}$ based on this. For example, we can use the maximum value of \eqref
{eq:10} over
all maximal ancillaries to assess whether or not $\Pi_{2}$ is weakly
informative relative to $\Pi_{1}$. When this maximum is small, we conclude\vadjust{\goodbreak}
that we have a small prior probability of finding evidence against the null
hypothesis of no prior-data conflict when using $\Pi_{2}$. We illustrate
this via an example.

%f6 ###
\begin{figure}

\includegraphics{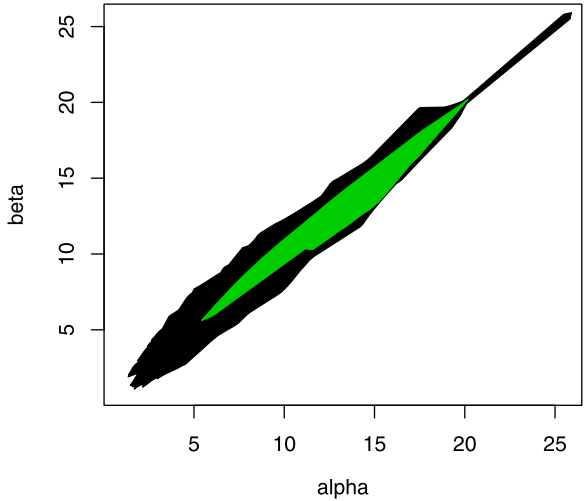}

  \caption{Plot of all $(\protect\alpha,\protect\beta)$ corresponding to
$\operatorname{Beta}(\alpha,\beta)$ priors that are weakly
informative at level $\protect\gamma=0.05$ (light and dark shading)
and uniformly weakly informative (light shading).}\label{fig6}
\end{figure}

\begin{example}
Suppose that we have a sample of~$n$ from the $\operatorname{Multinomial}(
1,(1-\theta)/6,(1+\theta)/6,(2-\theta)/6,(2+\theta)/6)$ distribution
where $\theta\in[ -1,1] $ is unknown. Then the counts
$(
f_{1},f_{2},f_{3},f_{4}) $ constitute a~mi\-nimal sufficient statistic
and $U_{1}=(f_{1}+f_{2}, f_{3}+f_{4})$ is ancillary, as is $%
U_{2}=( f_{1}+f_{4},f_{2}+f_{3}) $. Then
$T=(f_{1},f_{2},f_{3},f_{4}) | U_{1}$
is given by $f_{1} |\ U_{1}\sim\ \mathrm{Binomial}%
(f_{1}+f_{2},(1-\theta)/2)$ independent of $f_{3} |\ U_{1}\sim\
\mathrm{%
Binomial}(f_{3}+f_{4},(2-\theta)/4)$, giving
\begin{eqnarray*}
&&m_{T}(f_{1},f_{2},f_{3},f_{4} |
U_{1})\\
&&\quad=\pmatrix{{f_{1}+f_{2}}\cr{f_{1}}}{\pmatrix{f_{3}+f_{4}\cr{f_{3}}}}\\
&&\qquad\cdot\int_{-1}^{1}\biggl( \frac{1-\theta}{2}\biggr) ^{f_{1}}\biggl( \frac{%
1+\theta}{2}\biggr) ^{f_{2}}\biggl( \frac{2-\theta}{4}\biggr)
^{f_{3}}\\
&&\qquad{}\cdot\biggl( \frac{2+\theta}{4}\biggr) ^{f_{4}}\pi( \theta
)
\,d\theta.
\end{eqnarray*}
We then have two 1-dimensional distributions $f_{1} |
U_{1}$ and
$f_{3} \vert U_{1}$ to use for checking for prior-data conflict. A similar
result holds for the conditional distribution given $U_{2}$.

For example, suppose $\pi$ is a $\operatorname{Beta}(20,20)$ distribution on $[-1,1] $,
so the prior concentrates about~$0$, and for a sample of $n=18$
we have that $U_{1}=f_{1}+f_{2}=10$ and $U_{2}=f_{1}+f_{4}=8$. In
Figure \ref{fig6}
we have plotted all the values of $(\alpha,\beta)$ that correspond to
a~$\operatorname{Beta}(\alpha,\beta)$ prior that is weakly informative relative to the
$\operatorname{Beta}%
(20,20)$ prior at level $\gamma=0.05$, as well as those that are uniformly
weakly informative. So for each such $(\alpha,\beta)$ we have that
\eqref{eq:10} is
less than or equal to 0.05 for both $U=U_{1}$ and $U=U_{2}$.
\end{example}

%s6 ###
\vspace*{3pt}\section{Conclusions}\vspace*{3pt}\label{sec:06}

We have developed an approach to measuring the amount of information a prior
puts into a statistical analysis relative to another base prior. This base
prior can be considered as the prior that best reflects current information
and our goal is to determine a~prior that is weakly informative with respect
to it. Our measure is in terms of the prior predictive probability, using
the base prior, of obtaining a prior-data conflict. This was applied in
several examples where the approach is seen to give intuitively reasonable
results. The examples chosen here focused on commonly used prior
families. In several
cases these were conjugate families, although there is no special advantage
computationally to conjugacy in this context.

As noted in several examples, we need to be careful when we conceive of a
prior being weakly informative relative to another. Ultimately this concept
needs to be made precise and we feel our definition is a reasonable
proposal. The definition has intuitive support, in terms of avoiding
prior-data conflicts, and provides a quantifiable criterion that can be used
to select priors.

In any application we should still check for prior-data conflict for the
base prior using \eqref{eq:03}. If prior-data conflict is found, a
substitute prior
that is weak\-ly informative relative to the base prior can then be
selected and a check made for prior-data conflict with respect to the new
prior.
While selecting the prior based on the observed data is not ideal, this process
at least seems defensible from a logical perspective. For example,
the new prior still incorporates some of the information
from the base prior and is not entirely driven by the data. Certainly,
in the end it seems preferable to base an analysis on a~prior for which
a prior-data conflict does not exist.
Of course, we must still report the original conflict and how this was resolved.

We have restricted our discussion here to proper priors.
The concept of weak informativity is obviously related to the idea of
noninformativity and improper priors.
Certainly any prior that has a claim to being noninformative
should not lead to prior-data conflict.
At this time, however, there is no precise definition of what a
noninformative prior is,
whereas we have provided a definition of a weakly informative prior. In
the examples
of Section \ref{sec:03.1} and \ref{sec:03.2} we see that if the spread
of $\Pi_2$ is made large enough, then $\Pi_2$ is uniformly
weakly informative with respect to the base prior. This suggests that
the flat improper
prior, which is Jeffreys' prior for this problem, can be thought of as
always being uniformly weakly informative.
The logistic regression example of Section \ref{sec:04.3} suggests
caution, however, in interpreting increased diffuseness
as a characterization of weak informativity.
In the binomial example of Section \ref{sec:04.1}
the uniform prior is always weakly informative with respect to the base prior,
while the $\operatorname{Beta}(1/2,1/2)$ (Jeffreys') prior is not.
Further work
is required for a full examination of the
relationships among the concepts of prior-data conflict,
noninformativity and weak informativity.

\begin{appendix}\label{app}
\section*{Appendix}

\begin{pf*}{Proof of Lemma \protect\ref{lem:01}}
We have that $x_{\gamma}=\gamma$ since $P_{1}(t)$ has a continuous
distribution under $M_{1T}$. Suppose $m_{iT}^{\ast}(t)$ has a point
mass at
$r_{0}$ when $t\sim M_{iT}$. The assumption $M_{iT}(m_{iT}^{\ast
}(t)=r_{0})>0$ implies\break $(m_{iT}^{\ast})^{-1}\{r_{0}\}\not=\varnothing$.
Then, pick $t_{r_{0}}\in(m_{iT}^{\ast})^{-1}\{r_{0}\}$ so that $%
m_{iT}^{\ast}(t_{r_{0}})=r_{0}$ and let $\eta_{i}=P_{i}(t_{r_{0}})$. Then,
$P_{i}(t)$ has point mass at $\eta_{i}$ because $M_{iT}(P_{i}(t)=\eta
_{i})\geq\break M_{iT}(m_{iT}^{\ast}(t)=m_{iT}^{\ast}(t_{r_{0}}))=M_{iT}(m_{iT}^{\ast}(t)=r_{0})>0$.\break
This is a contradiction and so $m_{iT}^{\ast}(t)$ has a continuous distribution when $t\sim M_{iT}$.

Let $r_{\gamma}=\sup\{r\in\mathcal{R}\dvtx M_{2T}(m_{2T}^{\ast}(t)\leq
r)\leq\gamma\}$ where $\mathcal{R}=\overline{\{m_{2T}^{\ast}(t)\dvtx t\in
\mathcal{T}\}}$ and $\mathcal{T}$ is the range space of~$T$. Then,
$M_{2T}(m_{2T}^{\ast}(t)\leq r_{\gamma}) = \gamma$ and
$M_{2T}(m_{2T}^{*}(t) \leq\break r_{\gamma}+\epsilon)>\gamma$
for all $\epsilon>0$. Thus, we have that $\{t\dvtx\break  P_{2}(t)\leq\gamma\}=\{t\dvtx m_{2T}^{\ast}(t)\leq r_{\gamma}\}$,
$M_{1T}(P_{2}(t)\leq\gamma)=\break M_{1T}(m_{2T}^{\ast}(t)\leq r_{\gamma})$, and
$\Pi_{2}$ is weakly informative at level $\gamma$ relative to $\Pi_{1}$
if and only if $M_{1T}(m_{2T}^{\ast}(t)\leq r_{\gamma})\leq\gamma$. The
fact that $\{r_{\gamma}\dvtx \gamma\in\lbrack0,1]\}\subset\mathcal{R}$
implies the last statement.
\end{pf*}

\begin{pf*}{Proof of Theorem \protect\ref{thm:01}}
Suppose first that $\Sigma_{1}\leq\Sigma_{2}$. We have that $%
n^{-1}I+\Sigma_{1}\leq n^{-1}I+\Sigma_{2}$ and so $(n^{-1}I+\Sigma
_{1})^{-1}\geq(n^{-1}I+\Sigma_{2})^{-1}$. This implies that~\eqref
{eq:07} is less
than $\gamma$ and so the $N_{k}(\mu_{0},\Sigma_{2})$ prior is uniform\-ly
weakly informative relative to the $N_{k}(\mu_{0},\Sigma_{1})$ prior.

For the converse put $V_{i}=\{y\dvtx y^{\prime}(n^{-1}I+\Sigma
_{i})^{-1}y\leq
1\}$. If $V_{1}\subset V_{2}$, then for $y\in R^{k}\backslash\{0\}$ there
exists $c>0$ such that $c^{2}y^{\prime}(n^{-1}I+\Sigma_{1})^{-1}y=1$ which
implies $cy\in V_{2}$ and so $c^{2}y^{\prime}(n^{-1}I+\Sigma
_{2})^{-1}y\leq1$. This implies that $y^{\prime}(n^{-1}I+\Sigma
_{1})^{-1}y\geq y^{\prime}(n^{-1}I+\Sigma_{2})^{-1}y$ and so $\Sigma
_{1}\leq\Sigma_{2}$ and the result follows. If $V_{2}\subset V_{1}$, then
the same reasoning says that $\Sigma_{2}\leq\Sigma_{1}$ and \eqref
{eq:07} would be
greater than $\gamma$ if $\Sigma_{2}<\Sigma_{1}$.

So we need only consider the case where $V_{1}\cap V_{2}^{c}$,
$V_{1}^{c}\cap V_{2}$ both have positive volumes, that is, we are
supposing that
neither $\Sigma_{2}-\Sigma_{1}$ nor $\Sigma_{1}-\Sigma_{2}$ is positive
semidefinite and then will obtain a contradiction. Let $\delta=\inf
\{y^{\prime}(n^{-1}I+\Sigma_{1})^{-1}y\dvtx y\in V_{1}\cap\partial V_{2}\}$
and note that $\delta<1$, since $V_{1}^{o}\cap\partial V_{2}\neq\phi$,
that is, there are points in the interior of $V_{1}$\ on the
\mbox{boundary}
of~$V_{2}$. Now put $V_{0}=\{y\in V_{1}\cap V_{2}^{c}\dvtx y^{\prime}(n^{-1}I+\Sigma
_{1})^{-1}y\leq(1+\delta)/2\}$ and note that $V_{0}$ has positive volume.

Let $Y\sim N_{k}(0,n^{-1}I+\Sigma_{1})$ and $\tau_{\gamma
}^{2}=G_{k}^{-1}(1-\gamma)$.\vspace*{1pt} Then $M_{1T}(P_{1}(t)\,{\leq}\,\gamma\vspace*{1pt}
)\,{=}\,P(Y^{\prime}(n^{-1}I+\Sigma_{1})^{-1}Y\,{\geq}\,\tau_{\gamma}^{2})=P(Y\notin\tau
_{\gamma}V_{1})=1-P_{Y}(\tau_{\gamma}(V_{1}\cap V_{2})\cup\tau
_{\gamma}(V_{1}\cap V_{2}^{c}))$\vspace*{1pt} while $M_{1T}(P_{2}(t)\leq\gamma)=P(Y^{\prime
}(n^{-1}I+\Sigma_{2})^{-1}\cdot Y\geq\vspace*{1pt}\tau_{\gamma}^{2})=P(Y\notin\tau
_{\gamma}V_{2})=1-P_{Y}(\tau_{\gamma}(V_{1}\cap V_{2})\cup\tau
_{\gamma}(V_{1}^{c}\cap V_{2}))$.\vspace*{1pt} Since $\gamma=M_{1T}(P_{1}(t)\leq\gamma)$,\vspace*{1pt} we
need only show that $P_{Y}(\tau_{\gamma}(V_{1}\cap
V_{2}^{c}))>P_{Y}(\tau
_{\gamma}(V_{1}^{c}\cap V_{2}))$ for all $\gamma$ sufficiently small, to
establish the result.

Let $f(x)=k_{1}e^{-x/2}$ be such that\vspace*{1pt} $f(y^{\prime}(n^{-1}I+\break\Sigma
_{1})^{-1}y)$ is the density of $Y$. Then\vspace*{1pt} $P_{Y}(\tau_{\gamma
}(V_{1}^{c}\cap V_{2}))=\vspace*{1pt}\int_{\tau_{\gamma}(V_{1}^{c}\cap
V_{2})}f(y^{\prime}(n^{-1}I+\Sigma_{1})^{-1}y)\, dy\leq f(\tau_{\gamma
}^{2}y_{\ast}^{\prime}(n^{-1}I+\Sigma_{1})^{-1}y_{\ast}) \operatorname{Vol}
((V_{1}^{c}\cap V_{2}))\tau_{\gamma}^{k}$ where $y_{\ast}=\arg\min
\{y^{\prime}\cdot (n^{-1} I+\Sigma_{1})^{-1}y\dvtx y\in V_{1}^{c}\cap V_{2}\}$. Note\vspace*{1pt}
it is clear that $y_{\ast}\in\partial V_{1}$ and so $y_{\ast}^{\prime
}(n^{-1}I+\Sigma_{1})^{-1}y_{\ast}=1$ and $f(\tau_{\gamma
}^{2}y_{\ast
}^{\prime}\cdot (n^{-1}I+\Sigma_{1})^{-1}y_{\ast})=k_{1}e^{-\tau_{\gamma
}^{2}/2}$. Also,\vspace*{1pt} $P_{Y}(\tau_{\gamma}(V_{1}\cap V_{2}^{c}))\geq
P_{Y}(\tau
_{\gamma}V_{0})=\int_{\tau_{\gamma}V_{0}}f(y^{\prime}(n^{-1}I+\Sigma
_{1})^{-1}y)\, dy\geq f(\tau_{\gamma}^{2}(1+\delta)/2)
\operatorname{Vol}(V_{0})\tau
_{\gamma}^{k}$ where $f(\tau_{\gamma}^{2}(1+\delta)/2)=k_{1}e^{-\tau
_{\gamma}^{2}(1\vspace*{1pt}+\delta)/4}$. Therefore, as $\gamma\rightarrow0$,%
\begin{eqnarray*}
&&\frac{P_{Y}(\tau_{\gamma}(V_{1}\cap V_{2}^{c}))}{P_{Y}(\tau_{\gamma
}(V_{1}^{c}\cap V_{2}))}\\
&&\quad\geq e^{\tau_{\gamma}^{2}(1-\delta)/4}\frac
{\operatorname{%
Vol}((V_{1}^{c}\cap V_{2}))}{\operatorname{Vol}(V_{0})}\rightarrow\infty,
\end{eqnarray*}
since $\tau_{\gamma}=(G_{k}^{-1}(1-\gamma))^{1/2}\rightarrow\infty
$ as $%
\gamma\rightarrow0$ and $0<\delta<1$.
\end{pf*}

\begin{pf*}{Proof of Theorem \protect\ref{thm:02}}
First note that we can use~\eqref{eq:02} instead of \eqref{eq:03} in
this case as $J_{T}(x)$ is
constant in this case. We assume without loss of generality that $\mu
_{0}=0. $

We first establish several useful technical results. If $\Pi_{i}$ is a
probability distribution that is unimodal and symmetric about 0, and
$\phi
_{\nu}$ denotes a $N(0,\nu)$ density, we have that $m_{iT}(t)=\int
_{R}\phi
_{\nu}(t-\mu) \Pi_{i}(d\mu)$ is unimodal and symmetric about 0. We have
the following result.%\smallskip

\begin{alemma}\label{lem:a1}
If $T$ is a minimal sufficient \mbox{statistic}, $%
J_{T}(x)$ is constant in $x$, $\Pi_{1}$ and $\Pi_{2}$ are unimodal and
symmetric about 0, the $P_{i}(t)$ have continuous distributions when
$t\sim
M_{iT},m_{1T}(0)>m_{2T}(0)$, and $m_{1T}(t)=m_{2T}(t)$ has a unique solution
for $t>0$$, \ $then $\Pi_{2}$ is uniformly weakly informative relative
to $\Pi_{1}$.%\smallskip
\end{alemma}

\begin{pf}
By the unimodality and symmetry\break of~$m_{iT}$, we must
have that $P_{i}(t)=M_{iT}(m_{iT}(u)\leq m_{iT}(t))=M_{iT}(|u|\geq|t|)$.
We show $M_{1T}(|t|\geq t_{0})\leq M_{2T}(|t|\geq t_{0})$ for all $t_{0}>0$
because it is equivalent to $\Pi_2$ being uniformly weakly informative
relative to~$\Pi_1$ by Lemma \ref{lem:01}.
Let $t_{s}$ be the
solution of $m_{1T}(t)=m_{2T}(t)$ on $(0,\infty)$. From the unique solution
assumption, $m_{1T}(t)>m_{2T}(t)$ for $t\in(0,t_{s})$ and $%
m_{1T}(t)<m_{2T}(t)$ for $t>t_{s}$. For $0\leq t_{0}<t_{s},
M_{1T}(|t|\geq
t_{0})=2\int_{t_{0}}^{\infty}m_{1T}(t) \,dt=1-2\int
_{0}^{t_{0}}m_{1T}(t) \,dt\leq
1-\break2\int_{0}^{t_{0}}m_{2T}(t) \,dt=2\int_{t_{0}}^{\infty
}m_{2T}(t) \,dt=M_{2T}(|t|\geq t_{0})$ and for $t_{0}\geq t_{s}$,
$M_{1T}(|t|\geq t_{0})=2\int_{t_{0}}^{\infty}m_{1T}(t) \,dt\leq\break
2\int_{t_{0}}^{\infty}m_{2T}(t) \,dt,{=}\,M_{2T}(|t|\,{\geq}\, t_{0})$.
Thus, we are done.$\!\!\!$~%\smallskip
\end{pf}

We can apply Lemma \ref{lem:a1} to comparing normal and~$t$ priors when
sampling from
a normal.%\smallskip

\begin{alemma}\label{lem:a2}
Suppose we have a sample of $n$ from a location
normal model, $\Pi_{1}$ is a $N(0,\sigma_{1}^{2})$ prior and $\Pi
_{2}$ is
a $t_{1}(0,\sigma_{2}^{2},\lambda)$ prior. If $m_{1T}(0)>m_{2T}(0)$,
then $%
\Pi_{2}$ is uniformly weakly informative relative to~$\Pi_{1}$.%
\end{alemma}

\begin{pf}
We have that $m_{1T}=\phi_{1/n+\sigma_{1}^{2}}$
and, using the representation of the $t(\lambda)$ distribution as a gamma
mixture of normals, we write $m_{2T}(t)= \int_{0}^{\infty}\phi
_{1/n+\sigma_{2}^{2}/u}(t)k_{\lambda}(u) \,du$
where $k_{\lambda}$ is the
density of $\operatorname{Gamma}_{\mathrm{rate}}(\lambda/2,\lambda/2)$
distribution.
By the
symmetry of $\phi_{v}$, $m_{2T}$ is symmetric. Also, $\phi
_{v}(t_{1})>\phi_{v}(t_{2})$ for $0\,{\leq}\,t_{1}\,{<}\,t_{2}$
and so $m_{2T}(t_{1})\,{=}\,\int\!\phi
_{1/n+\sigma_{2}^{2}/u}(t_{1})k_{\lambda}(u) \,du\geq\int\phi
_{1/n+\sigma_{2}^{2}/u}(t_{2})k_{\lambda}(u) \,du=m_{2T}(t_{2})$.
Thus, $m_{2T}$ is decreasing on $(0,\infty)$, that is, $m_{2T}$ is
unimodal. To show
that $m_{2T}(t)$ is log-convex with respect to $t^{2}$, we prove that %
$(d^{2}/d(t^{2})^{2}) \log m_{2T}(t) \geq0$. Note that\vspace*{2pt}
$(d/d(t^{2}))\phi
_{v}(t) = (d/d(t^{2}))[(2\pi v)^{-1/2}\exp\{-t^{2}/ 2v\}] =-\phi_{v}(t)/2v$,
\begin{eqnarray*}
&&\frac{dm_{2T}(t)}{dt^{2}}\\
&&\quad= -\int_{0}^{\infty}\frac{\phi_{1/n+\sigma
_{2}^{2}/u}(t)}{2(1/n+\sigma_{2}^{2}/u)}k_{\lambda}(u) \,du, \\
&&\frac{d^{2}\log m_{2T}(t)}{d(t^{2})^{2}}\\
&&\quad= \frac{1}{m_{2T}(t)} \int_{0}^{\infty}\frac{\phi_{1/n+\sigma
_{2}^{2}/u}(t)}{[2(1/n+\sigma
_{2}^{2}/u)]^{2}}k_{\lambda}(u) \,du\\
&&\qquad{}-\frac{1}{m_{2T}(t)^{2}}\biggl(\int_{0}^{\infty}\frac{\phi
_{1/n+\sigma
_{2}^{2}/u}(t)}{2(1/n+\sigma_{2}^{2}/u)}k_{\lambda}(u) \,du\biggr)^{2}
\end{eqnarray*}
and so $d^{2}\log m_{2T}(t)/d(t^{2})^{2}= \operatorname{Var}_{V}([2(1/n+\break\sigma
_{2}^{2}/ V)]^{-1})\geq0$, where $V$ is the random variable having
density $%
\phi_{1/n+\sigma_{2}^{2}/v}(t)k_{\lambda}(v)/m_{2T}(t)$. Thus, $m_{2T}(t)$
is log-convex in $t^{2}$.

The functions $m_{1T}(t)$ and $m_{2T}(t)$ meet in at most two points on
$%
(0,\infty)$ because $\log m_{1T}(t)$ is linear in $t^{2}$ and $\log
m_{2T}(t)$ is convex in $t^{2}$. Also, $m_{1T}(t)$ and $m_{2T}(t)$
share at
least one point on $(0,\infty)$ because $m_{1T}(0)>m_{2T}(0)$, and the
following shows that $m_{1T}(t)<m_{2T}(t)$ for all large $t$. Note first
that if $u\geq\sigma_{2}^{2}/2\sigma_{1}^{2}$, then $(1/n+\sigma
_{1}^{2})/(1/n+\sigma_{2}^{2}/u)\geq1/2$ and $t^{2}/(u/n+\sigma
_{2}^{2})\geq(2\sigma_{1}^{2}/\sigma_{2}^{2})t^{2}/(1/n+2\sigma
_{1}^{2})$. Then,
\begin{eqnarray*}
&&\frac{m_{2T}(t)}{m_{1T}(t)} \\
&&\quad\geq\int_{\sigma_{2}^{2}/2\sigma
_{1}^{2}}^{\infty}\frac{(\lambda/2)^{\lambda/2}}{\Gamma(\lambda/2)}
\frac{(2\pi(1/n+\sigma_{2}^{2}/u))^{-1/2}}{(2\pi(1/n+\sigma
_{1}^{2}))^{-1/2}}u^{\lambda/2-1}\\
&&\hspace*{32pt}\qquad{}\cdot\frac{\exp\{-(u/2)(\lambda+t^{2}/({u/n+\sigma
_{2}^{2}}))\}%
}{\exp\{-(1/2)t^{2}/(1/n+\sigma_{1}^{2})\}} \,du \\
&&\quad\geq\frac{(\lambda/2)^{\lambda/2}}{\Gamma(\lambda/2)}\frac
{1}{2^{1/2}}%
\biggl(\frac{\sigma_{2}^{2}}{2\sigma_{1}^{2}}\biggr)^{\lambda
/2-1}\\
&&\qquad{}\cdot\int_{\sigma_{2}^{2}/2\sigma_{1}^{2}}^{\infty}
{\exp
\biggl\{-\biggl(\frac{u}{2}\biggr)\biggl(\lambda+\frac{(2\sigma_{1}^{2}/\sigma_{2}^{2})t^{2}}{1/n+2\sigma
_{1}^{2}}\biggr)\biggr\}}\\
&&\hspace*{38pt}\qquad{}\cdot\biggl(\exp\biggl\{\frac{-(1/2)t^{2}}{(1/n+\sigma_{1}^{2})}\biggr\}\biggr)^{-1} \,du \\
&&\quad=\frac{(\lambda/2)^{\lambda/2}}{\Gamma(\lambda/2)}\frac{1}{2^{1/2}}
\\
&&\qquad{}\cdot\biggl(\frac{\sigma_{2}^{2}}{2\sigma_{1}^{2}}\biggr)^{\lambda/2-1}\exp
\{-(1/2)(\sigma_{2}^{2}/2\sigma_{1}^{2})\lambda\}\\
&&\qquad{}\cdot{\exp\biggl\{\!\biggl(\frac{1}{2}\biggr)t^{2}\biggl(\!\biggl(\frac{1}{n}+\sigma
_{1}^{2}\biggr)^{-1}-\biggl(\frac{1}{n}+2\sigma_{1}^{2}\biggr)^{-1}\biggr)\!\biggr\}}\\
&&\qquad{}\cdot\biggl\{2^{-1}\biggl(\lambda+\frac{(2\sigma
_{1}^{2}/\sigma_{2}^{2})t^{2}}{(1/n+2\sigma_{1}^{2})}\biggr)\biggr\}^{-1}\rightarrow
\infty
\end{eqnarray*}
as $t^{2}\rightarrow\infty$.

The above conditions together imply that $m_{1T}(t)$ and $m_{2T}(t)$
meet in
exactly one point on $(0,\infty)$.\ Therefore, $\Pi_{2}$ is uniformly
weakly informative relative to $\Pi_{1}$ by Lemma \ref{lem:a1}.%
\end{pf}

Since $\int_{0}^{\infty}(1/n+\sigma^{2}/u)^{-1/2}k_{\lambda}(u) \,du$ is
strictly decreasing in $\sigma^{2}$, we see that $m_{1T}(0)=(2\pi
(1/n+\break\sigma_{1}^{2}))^{-1/2}\geq m_{2T}(0) =(2\pi)^{-1/2}\int
_{0}^{\infty
}(1/n+\sigma_{2}^{2}/u)^{-1/2}\cdot k_{\lambda}(u) \,du$ is equivalent to
$\sigma
_{2}\geq\sigma_{0n}$ where $\sigma_{0n}$ satisfies $(1/n+\sigma
_{1}^{2})^{-1/2}=\int_{0}^{\infty}(1/n+\sigma
_{0n}^{2}/u)^{-1/2}k_{\lambda
}(u) \,du$. This proves the first part of
Theorem \ref{thm:02}.

We also need the following results for the remaining parts of Theorem
\ref{thm:02}.%\smallskip

\begin{alemma}\label{lem:a3}
(\textup{i}) $\sigma_{0n}^{2}/\sigma_{1}^{2}$
increases as $n\sigma_{1}^{2}\rightarrow\infty$, (\textup{ii}) $\sigma
_{0n}^{2}/\sigma_{1}^{2}\rightarrow(2/\lambda)\Gamma^{2}((\lambda
+1)/2)/\Gamma^{2}(\lambda/2)$ as\break $n\sigma_{1}^{2}\rightarrow\infty
$.%\smallskip
\end{alemma}

\begin{pf}
{(i)} We have $n^{-1/2}(1/n+\sigma
_{1}^{2})^{-1/2}=\break n^{-1/2}\int_{0}^{\infty}(1/n+\sigma
_{0n}^{2}/u)^{-1/2}k_{\lambda}(u) \,du$ and putting $\alpha=n\sigma
_{1}^{2},\beta=n\sigma_{0n}^{2}$, we can write this as
%
%e11 ###
\setcounter{equation}{0}
\begin{equation}\label{eq:A1}
\qquad(1+\alpha)^{-1/2}=\int_{0}^{\infty}(1+\beta/u)^{-1/2}k_{\lambda
}(u) \,du.
\end{equation}
Differentiating both sides of \eqref{eq:A1} with respect to $\alpha$,
we have $(1+\alpha)^{-3/2}=\int_{0}^{\infty}(1+\beta/\break u)^{-3/2}u^{-1}k_{\lambda
}(u) \,du  (d\beta/d\alpha)$. If we let $U\sim\break\operatorname
{Gamma}_{\mathrm{rate}%
}(\lambda/2,\lambda/2)$, then this integral can be written as the
expectation
\begin{eqnarray*}
&& E\bigl((1+\beta/U)^{-3/2}U^{-1}\bigr)\\
&&\quad=E\bigl((1+\beta/U)^{-3/2}(\beta
/U+1-1)/\beta\bigr) \\
&&\quad =\beta^{-1}E\bigl((1+\beta/U)^{-1/2}\bigr)\\
&&\qquad{}-\beta^{-1}E\bigl((1+\beta/U)^{-3/2}\bigr)
\\
&&\quad \leq\beta^{-1}E\bigl((1+\beta/U)^{-1/2}\bigr)\\
&&\qquad{}-\beta^{-1}\bigl\{E\bigl((1+\beta
/U)^{-1/2}\bigr)\bigr\}^{3} \\
&&\quad=\beta^{-1}(1+\alpha)^{-1/2}-\beta^{-1}(1+\alpha)^{-3/2}\\
&&\quad=(1+\alpha
)^{-3/2}(\alpha/\beta),
\end{eqnarray*}
where the inequality follows via Jensen's inequality. Hence, $d\beta
/d\alpha=(1+\alpha)^{-3/2}[E((1+\beta/U)^{-3/2}\cdot U^{-1})]^{-1}\geq
\beta/\alpha$
and so $\beta/\alpha$ is an increasing function of $\alpha$ because $
d(\beta/\alpha)/d\alpha=\alpha^{-1}(d\beta/d\alpha)-\beta/\alpha
^{2}\geq0$.
This proves $\sigma_{0n}^{2}/\sigma_{1}^{2}=n\sigma_{0n}^{2}/n\sigma
_{1}^{2}=\beta/\alpha$ increases as $\alpha=n\sigma
_{1}^{2}\rightarrow
\infty$.

(ii) It is easy to check that $\beta=0$ when $%
\alpha=0$ and $\beta>0$ for $\alpha>0$. Let $\alpha_{0},\beta_{0}$
be a
pair satisfying $\alpha_{0}>0$ and \eqref{eq:A1}. Then, $\beta/\alpha
\geq\beta
_{0}/\alpha_{0}>0$ for $\alpha>\alpha_{0}$ and $\beta\rightarrow
\infty$
as $\alpha\rightarrow\infty$. Therefore,
\begin{eqnarray*}
\lim_{\alpha\rightarrow\infty}\biggl( \frac{\beta}{\alpha}\biggr)
^{1/2}& =&\lim_{\alpha\rightarrow\infty}\frac{\sqrt{\beta}}{\sqrt{%
1+\alpha}}\\
&=&\lim_{\alpha\rightarrow\infty}E\biggl( \frac{\sqrt{\beta
}}{%
\sqrt{1+\beta/U}}\biggr) \\
& =&\lim_{\beta\rightarrow\infty}E\biggl( \frac{\sqrt{\beta}}{\sqrt{%
1+\beta/U}}\biggr) =E\bigl(\sqrt{U}\bigr) \\
& =&\int_{0}^{\infty}\sqrt{u} k_{\lambda}(u) \,du\\
&=&(2/\lambda
)^{1/2}\Gamma
\bigl((\lambda+1)/2\bigr)/\Gamma(\lambda/2)
\end{eqnarray*}
and this proves (ii).%\smallskip
\end{pf}

\begin{alemma}\label{lem:a4}
Suppose we have a sample of $n$ from a location
normal model, $\Pi_{1}$ is a $N(0,\sigma_{1}^{2})$ prior and $\Pi
_{2}$ is
a $t_{1}(0,\sigma_{2}^{2},\lambda)$ prior. Then $\Pi_{2}$ is
asymptotically uniformly weakly informative relative to $\Pi_{1}$ if
and only if $\sigma_{2}^{2}/\sigma_{1}^{2}\geq(2/\lambda)\Gamma
^{2}((\lambda+1)/2)/\Gamma^{2}(\lambda/2)$.%\smallskip
\end{alemma}

\begin{pf}
Suppose that $\sigma_{2}^{2}/\sigma_{1}^{2}\geq
(2/\lambda)\Gamma^{2}((\lambda+1)/\break 2)/\Gamma^{2}(\lambda/2)$. Then by
Lemma \ref{lem:a3} $\sigma_{2}^{2}/\sigma_{1}^{2}\geq\sigma
_{0n}^{2}/\sigma
_{1}^{2}$ for all $n$ and so $\Pi_{2}$ is uniformly weakly informative with
respect to $\Pi_{1}$ for all $n$. So \eqref{eq:04} is bounded above by
$\gamma$ for
all $n$ and so the limiting value of \eqref{eq:04} is also bounded
above by $\gamma$.
This establishes that~$\Pi_{2}$ is asymptotically uniformly weakly
informative relative to $\Pi_{1}$.

Suppose now that $\sigma_{2}^{2}/\sigma_{1}^{2}<(2/\lambda)\Gamma
^{2}((\lambda+1)/ 2)/\break \Gamma^{2}(\lambda/2)$. Note that %
$m_{1T}(t) = \lim_{n\rightarrow\infty}m_{1T,n}(t)=\break(2\pi\sigma
_{1}^{2})^{-1/2}\exp(-t^{2}/(2\sigma_{1}^{2}))$ and $m_{2T}(t)=\break\lim_{n
\rightarrow\infty} m_{2T,n}(t) =\Gamma((\lambda+1)/2)/ (\Gamma
(\lambda/2)\sqrt{\pi\lambda\sigma
_{2}^{2}})(1+x^{2}/(\sigma_{2}^{2}\lambda))^{-(\lambda+1)/2}$.
Therefore, we get $m_{1T}(0)=\break  1/\sqrt{2\pi\sigma_{1}^{2}}<\Gamma
((\lambda+1)/2)/ \Gamma
(\lambda/ 2)\sqrt{\pi\lambda\sigma_{2}^{2}}=m_{2T}(0)$. Let $B=\{t\dvtx m_{2T}(t)> m_{1T}(0)\}$ and
$\gamma=M_{2T}(B^{c})$. Then, $%
m_{1T}(t)\leq m_{1T}(0)\leq m_{2T}(t)$ on $B$ and\break  $M_{1T}(P_{2}(t)\leq
\gamma)=M_{1T}(B^{c})=1-M_{1T}(B)=1-\break \int_{B}m_{1T}(t)\,dt\geq
1-\int_{B}m_{1T}(0)\,dt \geq1-\int_{B}m_{2T}(t)\,dt=M_{2T}(B^{c})=\gamma$.
Hence, $\Pi_{2}$ is not
weakly informative relative to $\Pi_{1}$\ at level $\gamma$.
Therefore, $\sigma_{2}^{2}/\sigma_{1}^{2}\geq(2/\break \lambda)\Gamma
^{2}((\lambda+1)/2)/\Gamma^{2}(\lambda/2)$.%\smallskip
\end{pf}

It is now immediate that $\sup_{\gamma\in\lbrack
0,1]}G_{1}^{-1}(1-\gamma
)/\break H_{1,\lambda}^{-1}(1-\gamma)=(2/\lambda)\Gamma^{2}((\lambda
+1)/2)/ \Gamma^{2}(\lambda/2)$ and the proof of Theorem \ref{thm:02}
is complete.
\end{pf*}

\begin{pf*}{Proof of Theorem \protect\ref{thm:03}}
Since the minimal sufficient statistic $T(x)=\bar{x}$ is linear, there
is no
volume distortion and we can use \eqref{eq:02} instead of \eqref
{eq:03}. The limiting prior
predictive distribution of $T(x)=\bar{x}$ under $\Pi_{1}$\ is $N(\mu
_{0},\Sigma_{1})$ and under $\Pi_{2}\ $it is\ $t_{k}(\mu_{0},\Sigma
_{2},\lambda)$. It is easy to check that $U_{1}=(T-\mu_{0})^{\prime
}\Sigma_{1}^{-1}(T-\mu_{0})\sim\chi^{2}(k)$ when $T\sim\Pi_{1}$
and $%
U_{2}=(T-\mu_{0})^{\prime}\Sigma_{2}^{-1}(T-\mu_{0})\sim
kF_{k,\lambda}$
when $T\sim\Pi_{2}$. This implies that $P_{2,n}(t_{0})$ converges to $
P_{2}(t_{0})=\Pi_{2}(\pi_{2}(t)\leq\pi_{2}(t_{0}))=1- H_{k,\lambda
}((t_{0}-\mu_{0})^{\prime}\Sigma_{2}^{-1}(t_{0}-\mu_{0})/k)$, where
$%
H_{k,\lambda}$ is the distribution function of an $F_{k,\lambda}$
distribution. Further, we have that \eqref{eq:04} converges to $\Pi
_{1}(P_{2}(t)\leq
\gamma)$.

Let $V_{i}=\{u\in R^{k}\dvtx u^{\prime}\Sigma_{i}^{-1}u<1\}$ for $i=1,2$. By
the continuity of $\Pi_{2}(\pi_{2}(t)\leq r)$ as a function of $r$, and
the continuity of $\pi_{2}(t)$, there exists $t_{0}$ such that $%
P_{2}(t)\leq\gamma$ if and only if $\pi_{2}(t)\leq\pi_{2}(t_{0})$.
Hence, $\Pi_{2}$ is asymptotically uniformly weakly informative
relative to
$\Pi_{1}$ if and only if $%
\Pi_{1}(\pi_{2}(t)\leq\pi_{2}(t_{0}))\leq\break\Pi_{2}(\pi_{2}(t)\leq
\pi_{2}(t_{0}))$ for all $t_{0}\in R^{k}$ by Lemma \ref{lem:01}. Since
$\pi_{2}(t)$ is decreasing in
$u_{2}=U_{2}(t)$, the set $\{\pi_{2}(t)\leq\pi
_{2}(t_{0})\}=\{u_{2}(t)\geq u_{2}(t_{0})\}=\mu_{0}+u_{2}(t_{0})V_{2}^{c}$.
So we must prove that $\Pi_{1}(\mu_{0}+r^{1/2}V_{2}^{c})\leq\Pi
_{2}(\mu
_{0}+r^{1/2}V_{2}^{c})$ for all $r\geq0$.

The positive semidefiniteness of $\Sigma_{2}-\tau_{\lambda}^{2}\Sigma
_{1} $ implies that $\Sigma_{1}^{-1}/\tau_{\lambda}^{2}-\Sigma_{2}^{-1}$
is positive semidefinite. Then, for $u\in V_{2}^{c}$, that is,
$u^{\prime
}\Sigma_{2}^{-1}u\geq1$, we have $u^{\prime}\Sigma_{1}^{-1}u=\tau
_{\lambda}^{2}\cdot u^{\prime}(\Sigma_{1}^{-1}/\tau_{\lambda
}^{2})u\geq
\tau_{\lambda}^{2}u^{\prime}\Sigma_{2}^{-1}u\geq\tau_{\lambda}^{2}$.
Thus, $V_{2}^{c}\subset\tau_{\lambda}V_{1}^{c}$.

Now we prove a stronger inequality $\Pi_{1}(\mu_{0}+\break r^{1/2}\tau
_{\lambda}\cdot V_{1}^{c})\leq\Pi_{2}(\mu_{0}+r^{1/2} V_{2}^{c})$ for all $r\geq0$. Note
that
\begin{eqnarray*}
&& \Pi_{1}(\mu_{0}+r^{1/2}\tau_{\lambda}V_{1}^{c})\\
&&\quad=\Pi
_{1}\bigl(u_{1}(t)\geq
r\tau_{\lambda}^{2}\bigr)\\
&&\quad=\int_{r\tau_{\lambda}^{2}}^{\infty}\frac
{2^{-k/2}}{%
\Gamma(k/2)}u^{k/2-1}e^{-u/2} \,du, \\
&& \Pi_{2}(\mu_{0}+r^{1/2}V_{2}^{c})\\
&&\quad=\Pi_{2}(u_{2}(t)\geq
r) \\
&&\quad=\int_{r/k}^{\infty}\frac{\Gamma((k+\lambda
)/2)}{\Gamma(k/2)\Gamma
(\lambda/2)}\\
&&\qquad{}\cdot\biggl(\frac{k}{\lambda}\biggr)^{k/2}u^{k/2-1}
\biggl(1+\frac{ku}{\lambda} \biggr)^{-(k+\lambda)/2} \,du,
\end{eqnarray*}
and set $f(r)=\Pi_{2}(\mu_{0}+r^{1/2}V_{2}^{c})-\Pi_{1}(\mu
_{0}+r^{1/2}\tau_{\lambda}\cdot V_{1}^{c})$.
Then, $f(0)=0$ and
\begin{eqnarray*}
\frac{df(r)}{dr}
&= &\frac{2^{-k/2}}{\Gamma(k/2)}(r\tau_{\lambda
}^{2})^{k/2-1}e^{-r\tau_{\lambda}^{2}/2}\tau_{\lambda}^{2}\\
&&{}\cdot\frac{\Gamma((k+\lambda)/2)}{\Gamma(k/2)\Gamma
(\lambda
/2)}\biggl(\frac{k}{\lambda}\biggr)^{k/2}\biggl(\frac{r}{k}
\biggr)^{k/2-1}\\
&&{}\cdot(1+r/\lambda)^{-(k+\lambda)/2}\frac{1}{k} \\
& =&\frac{(\tau_{\lambda}^{2}/2)^{k/2}}{\Gamma(k/2)}r^{k/2-1}e^{-r\tau
_{\lambda}^{2}/2}\\
&&{}-\frac{\Gamma((k+\lambda)/2)}{\Gamma(k/2)\Gamma
(\lambda/2)}\lambda^{-k/2}r^{k/2-1}\\
&&\quad{}\cdot(1+r/\lambda
)^{-(k+\lambda
)/2} \\
& =&p_{1}-p_{2}.
\end{eqnarray*}
Note that $p_{1}-p_{2}\geq0$ is equivalent to $p_{1}/p_{2}\geq1$. Further
recalling the definition of $\tau_{\lambda}^{2}$ from the statement
of the
theorem,
\begin{eqnarray*}
\frac{p_{1}}{p_{2}} &=& \frac{\tau_{\lambda}^{k}\Gamma(\lambda
/2)}{\Gamma
((k+\lambda)/2)}\biggl(\frac{\lambda}{2}\biggr)^{k/2}(1+r/\lambda
)%
^{(k+\lambda)/2}\\
&&{}\cdot\exp(-r\tau_{\lambda}^{2}/2) \\
&=& (1+r/\lambda)^{(k+\lambda)/2}\exp(-r\tau_{\lambda
}^{2}/2)\geq1.
\end{eqnarray*}
The logarithm of $p_{1}/p_{2}$ given by $\log(p_{1}/p_{2})=-r\tau
_{\lambda
}^{2}/\break2+((k+\lambda)/2)\log(1+r/\lambda)$ is concave as a function
of $%
r>0 $. Hence, $\log(p_{1}/p_{2})=0$ has exactly two solutions: $r=0$
and $%
r=r_{s}$. Because of its concavity, the function $\log(p_{1}/p_{2})$ is
positive on $(0,r_{s})$ and negative on $(r_{s},\infty)$. This implies that
$f(r)$ is increasing on $(0,r_{s})$ and decreasing on $(r_{s},\infty)$.
Since $f(0)=0$ and $\lim_{r\rightarrow\infty}f(r)=0$, the function
$f$ is
nonnegative, that is, $f(r)\geq0$ for all $r\geq0$. Thus, $\Pi
_{1}(\mu
_{0}+r^{1/2}V_{2}^{c})\leq\Pi_{1}(\mu_{0}+r^{1/2}\tau_{\lambda
}V_{1}^{c})\leq\Pi_{2}(\mu_{0}+r^{1/2}V_{2}^{c})$ for all $r\geq0$.
\end{pf*}

\begin{pf*}{Proof of Theorem \protect\ref{thm:04}}
Let $x_{c}^{-1}=\beta_{1}/(\alpha_{1}+\break1/2)=\beta_{2}/(\alpha_{2}+ 1/2)$.
For $i=1,2$, let $t_{i}(t_{0})=1/\break (x_{c}r_{i}(t_{0}))$ be the two solutions
of $m_{2T}^{\ast}(t_{i})=m_{2T}^{\ast}(t_{0})$ (one of the $t_{i}$
equals $%
t_{0})$ so $0<r_{1}\leq1\leq r_{2}$. Note that $r_{2}(t_{0})=1$ if and only
if $t_{0}=x_{c}^{-1}$ and then $r_{1}(t_{0})=1$ as well. Then, $\log
(r_{1}/r_{2})=r_{1}-r_{2}$ and $%
dr_{1}/dr_{2}=(r_{2}-1)r_{1}/[(r_{1}-1)r_{2}]$. Now $\{t\dvtx m_{2T}^{\ast
}(t)\leq m_{2T}^{\ast}(t_{0})\}=\{t\dvtx 1/t\leq x_{c}r_{1}(t_{0})\mbox{ or
}%
1/t\geq x_{c}r_{2}(t_{0})\}$. By Lemma \ref{lem:01} we have that
uniform weak
informativity is equivalent to $M_{1T}(m_{2T}^{\ast}(t)\leq
m_{2T}^{\ast
}(t_{0}))\leq M_{2T}(m_{2T}^{\ast}(t)\leq m_{2T}^{\ast}(t_{0}))$ for
all $%
t_{0}$ and so we must prove that $M_{1T}(t\notin
(t_{2}(t_{0}),t_{1}(t_{0})))=M_{1T}(1/t\leq x_{c}r_{1}(t_{0}) $ or $
1/t \geq\break
x_{c}r_{2}(t_{0}))=1-M_{1T}(x_{c}r_{1}(t_{0})\leq1/t\leq
x_{c}r_{2}(t_{0}))\leq1-M_{2T}(x_{c}r_{1}(t_{0})\leq1/t\leq
x_{c}r_{2}(t_{0}))$ for all $t_{0}$. Since $r_{1}$ is implicitly a function
of $r_{2}$, it is equivalent to prove that $M_{1T}(x_{c}r_{1}\leq
1/t\leq x_{c}r_{2})-M_{2T}(x_{c}r_{1}\leq\break1/t\leq x_{c}r_{2})\geq0$ for all $%
r_{2}\geq1$. Using $(r_{1}/r_{2})^{\alpha}=\break \exp(\alpha(r_{1}-r_{2}))$,
we have that the derivatives of the two terms are given by
\begin{eqnarray*}
p_{1}& =&\frac{d}{dr_{2}}\int_{x_{c}r_{1}}^{x_{c}r_{2}}c_{1}u^{\alpha
_{1}-1}e^{-\beta_{1}u}\,du\\
&=&c_{1}(x_{c}r_{2})^{\alpha_{1}-1}e^{-\beta
_{1}x_{c}r_{2}}x_{c}\\
&&{}-c_{1}(x_{c}r_{1})^{\alpha_{1}-1}e^{-\beta
_{1}x_{c}r_{1}}x_{c}%
\frac{dr_{1}}{dr_{2}} \\
& =&c_{1}x_{c}^{\alpha_{1}}r_{2}^{\alpha_{1}-1}e^{-\beta
_{1}x_{c}r_{2}}\\
&&{}\cdot\biggl(1-\frac{r_{2}-1}{r_{1}-1}\exp\bigl((r_{2}-r_{1})(\beta
_{1}x_{c}-\alpha
_{1})\bigr)\biggr), \\
p_{2}& =&c_{2}x_{c}^{\alpha_{2}}r_{2}^{\alpha_{2}-1}e^{-\beta
_{2}x_{c}r_{2}}\\
&&{}\cdot\biggl(1-\frac{r_{2}-1}{r_{1}-1}\exp\bigl((r_{2}-r_{1})(\beta
_{2}x_{c}-\alpha_{2})\bigr)\biggr),
\end{eqnarray*}
where $c_{i}=\beta_{i}^{\alpha_{i}}/\Gamma(\alpha_{i})$. Then, recalling
the definition of $x_{c}$, we have that the ratio %
$p_{1}/p_{2} = (c_{1}/c_{2})x_{c}^{\alpha_{1}-\alpha_{2}}\cdot r_{2}^{\alpha
_{1}-\alpha_{2}}e^{(\beta_{2}-\beta
_{1})x_{c}r_{2}} = (c_{1}/c_{2})x_{c}^{\alpha_{1}-\alpha
_{2}}(r_{2}e^{-r_{2}})^{\alpha_{1}-\alpha_{2}}$\break
strictly decreases as $r_{2}$ increases from $1$ to $\infty$ when
$\alpha_{1}>\alpha_{2}$
because $\alpha_{1}-\alpha_{2}=(\beta_{1}-\beta_{2})x_{c}\geq0$,
and is
identically 1 when $\alpha_{1}=\alpha_{2}$. Suppose then that $\alpha
_{1}>\alpha_{2}$ so there is at most one $r_{2}$ value where $p_{1}=p_{2}$
and the derivative is 0. If $(p_{1}/p_{2})|_{r_{2}=1}<1$, then $%
p_{1}-p_{2}<0 $ for all $r_{2}\geq1$ and $M_{1T}(x_{c}r_{1}\leq1/t\leq
x_{c}r_{2})-M_{2T}(x_{c}r_{1}\leq1/t\leq x_{c}r_{2})$ strictly decreases
from $0$. This cannot hold because $M_{1T}(x_{c}r_{1}\leq1/t\leq
x_{c}r_{2})-M_{2T}(x_{c}r_{1}\leq1/t\leq x_{c}r_{2})\rightarrow0$ as $
r_{2}\rightarrow\infty$. Hence, $(p_{1}/\break p_{2})|_{r_{2}=1}\geq1$ and $
M_{1T}(x_{c}r_{1}\leq1/t\leq x_{c}r_{2})-\break M_{2T}(x_{c}r_{1}\leq1/t\leq
x_{c}r_{2})$ increases from $0$ near $r_{2}=1$ and decreases to $0$ as $
r_{2}\rightarrow\infty$. Therefore, $M_{1T}(x_{c}r_{1}\leq1/t\leq
x_{c}r_{2})-M_{2T}(x_{c}r_{1}\leq1/t\leq x_{c}r_{2})$ goes up from $0$ and
down to $0$ as $r_{2}$ increases from $1$ to $\infty$, and we have $%
M_{1T}(x_{c}r_{1}\leq1/t\leq x_{c}r_{2})-M_{2T}(x_{c}r_{1}\leq1/t\leq
x_{c}r_{2})\geq0$ for all $r_{2}\geq1$.
\end{pf*}
\end{appendix}

\section*{Acknowledgments}
The authors thank the Editor, Associate Editor and referees for many
helpful comments.

% imsref loaded by akundreckaite, 2011-06-14 10:50:23
% imsref loaded by akundreckaite, 2011-06-14 10:50:39


\begin{thebibliography}{13}
% BibTex style file: ims.bst, 2011-05-30
% Default style options (sort=0,type=number).
% Used options (sort=1,type=nameyear).

%b1 ###
\bibitem[\protect\citeauthoryear{Bernardo}{1979}]{Bernardo1979}
\begin{barticle}[author]
\bauthor{\bsnm{Bernardo},~\bfnm{Jose-M.}\binits{J.-M.}}
(\byear{1979}).
\btitle{Reference posterior distributions for {B}ayesian inference (with discussion)}.
\bjournal{J. Roy. Statist. Soc. Ser. B}
\bvolume{41}
\bpages{113--147}.
\bid{mr={0547240}}
\end{barticle}
\endbibitem

%b2 ###
\bibitem[\protect\citeauthoryear{Chib and Ergashev}{2009}]{ChibErgashev2009}
\begin{barticle}[author]
\bauthor{\bsnm{Chib},~\bfnm{Siddartha}\binits{S.}} \AND
  \bauthor{\bsnm{Ergashev},~\bfnm{Bakhodir}\binits{B.}}
(\byear{2009}).
\btitle{Analysis of multifactor affine yield curve models}.
\bjournal{J. Amer. Statist. Assoc.}
\bvolume{104}
\bpages{1324--1337}.
\end{barticle}
\endbibitem\

%b3 ###
\bibitem[\protect\citeauthoryear{Evans and Jang}{2010}]{EvansJang2010annstat}
\begin{barticle}[author]
\bauthor{\bsnm{Evans},~\bfnm{Michael}\binits{M.}} \AND
  \bauthor{\bsnm{Jang},~\bfnm{Gun~Ho}\binits{G.~H.}}
(\byear{2010}).
\btitle{Invariant {$P$}-values for model checking}.
\bjournal{Ann. Statist.}
\bvolume{38}
\bpages{312--323}.
\bid{mr={2589329}}
\end{barticle}
\endbibitem

%b4 ###
\bibitem[\protect\citeauthoryear{Evans and Jang}{2011}]{EvansJang2010b}
\begin{barticle}[author]
\bauthor{\bsnm{Evans},~\bfnm{Michael}\binits{M.}} \AND
  \bauthor{\bsnm{Jang},~\bfnm{Gun~Ho}\binits{G.~H.}}
(\byear{2011}).
\btitle{A limit result for the prior predictive
applied to checking for prior-data conflict}.
\bjournal{Statist. Probab. Lett.}
\bvolume{81}
\bpages{1034--1038}.
\end{barticle}
\endbibitem

%b5 ###
\bibitem[\protect\citeauthoryear{Evans and Moshonov}{2006}]{EvansMoshonov2006}
\begin{barticle}[author]
\bauthor{\bsnm{Evans},~\bfnm{Michael}\binits{M.}} \AND
  \bauthor{\bsnm{Moshonov},~\bfnm{Hadas}\binits{H.}}
(\byear{2006}).
\btitle{Checking for prior-data conflict}.
\bjournal{Bayesian Anal.}
\bvolume{1}
\bpages{893--914}.
\bid{mr={2282210}}
\end{barticle}
\endbibitem

%b6 ###
\bibitem[\protect\citeauthoryear{Evans and Moshonov}{2007}]{EvansMoshonov2007}
\begin{bincollection}[author]
\bauthor{\bsnm{Evans},~\bfnm{Michael}\binits{M.}} \AND
  \bauthor{\bsnm{Moshonov},~\bfnm{Hadas}\binits{H.}}
(\byear{2007}).
\btitle{Checking for prior-data conflict with hierarchically specified priors}.
In \bbooktitle{Bayesian Statistics and Its Applications}
(\beditor{\bfnm{A.~K.}\binits{A.~K.}~\bsnm{Upadhyay}},
  \beditor{\bfnm{U.}\binits{U.}~\bsnm{Singh}} \AND
  \beditor{\bfnm{D.}\binits{D.}~\bsnm{Dey}}, eds.)
\bpages{145--159}.
\bpublisher{Anamaya Publishers}, \baddress{New Delhi}.
\end{bincollection}
\endbibitem

%b7 ###
\bibitem[\protect\citeauthoryear{Gelman}{2006}]{Gelman2006}
\begin{barticle}[author]
\bauthor{\bsnm{Gelman},~\bfnm{Andrew}\binits{A.}}
(\byear{2006}).
\btitle{Prior distributions for variance parameters in hierarchical models}.
\bjournal{Bayesian Anal.}
\bvolume{1}
\bpages{515--533}.
\bid{mr={2221284}}
\end{barticle}
\endbibitem

%b8 ###
\bibitem[\protect\citeauthoryear{Gelman et~al.}{2008}]{Gelman-etal2008}
\begin{barticle}[author]
\bauthor{\bsnm{Gelman},~\bfnm{Andrew}\binits{A.}},
  \bauthor{\bsnm{Jakulin},~\bfnm{Aleks}\binits{A.}},
  \bauthor{\bsnm{Pittau},~\bfnm{Maria~Grazia}\binits{M.~G.}} \AND
  \bauthor{\bsnm{Su},~\bfnm{Yu-Sung}\binits{Y.-S.}}
(\byear{2008}).
\btitle{{A weakly informative default prior distribution for logistic and other
  regression models.}}
\bjournal{Ann. Appl. Statist.}
\bvolume{2}
\bpages{1360--1383}.
\bid{mr={2655663}}
\end{barticle}
\endbibitem

%b9 ###
\bibitem[\protect\citeauthoryear{Kass and Wasserman}{1995}]{KassWasserman1995}
\begin{barticle}[author]
\bauthor{\bsnm{Kass},~\bfnm{Robert~E.}\binits{R.~E.}} \AND
  \bauthor{\bsnm{Wasserman},~\bfnm{Larry}\binits{L.}}
(\byear{1995}).
\btitle{A reference {B}ayesian test for nested hypotheses and its relationship
  to the {S}chwarz criterion}.
\bjournal{J. Amer. Statist. Assoc.}
\bvolume{90}
\bpages{928--934}.
\bid{mr={1354008}}
\end{barticle}
\endbibitem

%b10 ###
\bibitem[\protect\citeauthoryear{Lindley}{1956}]{Lindley1956}
\begin{barticle}[author]
\bauthor{\bsnm{Lindley},~\bfnm{D.~V.}\binits{D.~V.}}
(\byear{1956}).
\btitle{On a measure of the information provided by an experiment}.
\bjournal{Ann. Math. Statist.}
\bvolume{27}
\bpages{986--1005}.
\bid{mr={0083936}}
\end{barticle}
\endbibitem

%b11 ###
\bibitem[\protect\citeauthoryear{Racine et~al.}{1986}]{Racine-etal1986}
\begin{barticle}[author]
\bauthor{\bsnm{Racine},~\bfnm{A.}\binits{A.}},
  \bauthor{\bsnm{Grieve},~\bfnm{A.~P.}\binits{A.~P.}},
  \bauthor{\bsnm{Fl{\"{u}}hler},~\bfnm{H.}\binits{H.}} \AND
  \bauthor{\bsnm{Smith},~\bfnm{A.~F.~M.}\binits{A.~F.~M.}}
(\byear{1986}).
\btitle{Bayesian methods in practice: Experiences in the pharmaceutical
  industry (with discussion)}.
\bjournal{J. Roy. Statist. Soc. Ser.~C}
\bvolume{35}
\bpages{93--150}.
\bid{mr={0868007}}
\end{barticle}
\endbibitem

%b12 ###
\bibitem[\protect\citeauthoryear{Tjur}{1974}]{Tjur1974}
\begin{bmisc}[author]
\bauthor{\bsnm{Tjur},~\bfnm{T.}\binits{T.}}
(\byear{1974}).
\bhowpublished{\textit{Conditional Probability Models}.
Institute of Mathematical Statistics, Univ. Copenhagen, Copenhagen.}
\bid{mr={0345151}}
\end{bmisc}
\endbibitem
\vspace*{-3pt}
\end{thebibliography}
\end{document}